# Practical Applications of Cosmology to Human Society

Eric J. Chaisson
Harvard-Smithsonian Center for Astrophysics
Harvard University, Cambridge, Massachusetts 02138 USA
www.cfa.harvard.edu/~ejchaisson

**Abstract**

Complex systems throughout Nature display structures and functions that are built and maintained, at least in part, by optimal energies flowing through them—not specific, ideal values, rather ranges in energy rate density below which systems are starved and above which systems are destroyed. Cosmic evolution, as a physical cosmology that notably includes life, is rich in empirical findings about many varied systems that can potentially help assess global problems facing us here on Earth. Despite its grand and ambitious objective to unify theoretical understanding of all known complex systems, cosmic evolution does have useful, practical applications from which humanity could benefit. Cosmic evolution's emphasis on quantitative data analyses might well inform our attitudes toward several serious issues now challenging 21$^{st}$-century society, including global warming, smart machines, world economics, and cancer research. This paper, which is a sequel to an earlier one that more fully articulates the expansive cosmic-evolutionary scenario from big bang to humankind, comprises one physicist's conjectures about each of these applied topics, suggesting how energy-flow modeling can guide our search for viable solutions to real-world predicaments confronting civilization today.

**Key Words:** evolution, complexity, energy, climate, machines, economics, cancer

## I. Introduction

How wonderful it would be if cosmic-evolutionary research prompted novel insights and practical applications for some of humankind's foremost challenges today. Such a natural-science survey, broadly addressing all known ordered systems across all of cosmic time, and potentially identifying a complexity metric of wide significance from quarks to quasars and from microbes to minds, does seemingly offer humanity some guidance at a time of accelerating global troubles on planet Earth.

This is a sequel to Paper I (Chaisson 2014), which summarized the full scenario of cosmic evolution as a scientific worldview that grants humans, as products of big history, a sense of place in the Universe. Yet, this highly interdisciplinary subject is more than an inclusive, subjective narrative of all that we witness in Nature; rather, as an objective study of change writ large, cosmic evolution is firmly grounded in natural science, in fact quantitatively so across many orders of magnitude in size, scale, time, and complexity. Nonetheless, its immense scope should not preclude specific, practical applications of real and useful merit for humanity and its vexatious society today.

Throughout the history of the Universe, as each type of ordered system became more complex, its normalized energy budget increased. Expressed as an energy rate density, $\Phi_m$, a hierarchical scheme ranks known organized structures that have experienced, in turn, physical, biological, and cultural evolution: stars and galaxies ($\Phi_m = 10^{-2}$-$10^2$ erg/s/g), plants and animals ($10^3$-$10^5$), society and machines ($\geq 10^5$). Figure 1 sketches the rise in complexity among Nature's many varied systems by plotting the change of energy rate density everywhen in time, from the beginning of the Universe to the present. Such a broad synthesis of natural science encapsulates the sum of "big history," demonstrating in a single graph the interconnectedness of galaxies, stars, planets, life, and society. This figure was discussed at length in Paper I (and more succinctly in Chaisson 2013), as was its core hypothesis that $\Phi_m$ is a complexity metric that compactly compares commonalities among increasingly complex systems throughout the natural sciences. Notably stressed in that earlier paper are various optimal energy ranges characterizing numerous complex systems—specifically, ranges in energy rate density that are empirically revealed by consistent, uniform analyses of a surprisingly wide spectrum of complex systems observed in Nature. This is cosmic evolution's iconic graph against which we examine how this grand subject might conceivably be of practical relevance, and even importance, to worldly issues now confronting humankind on Earth.

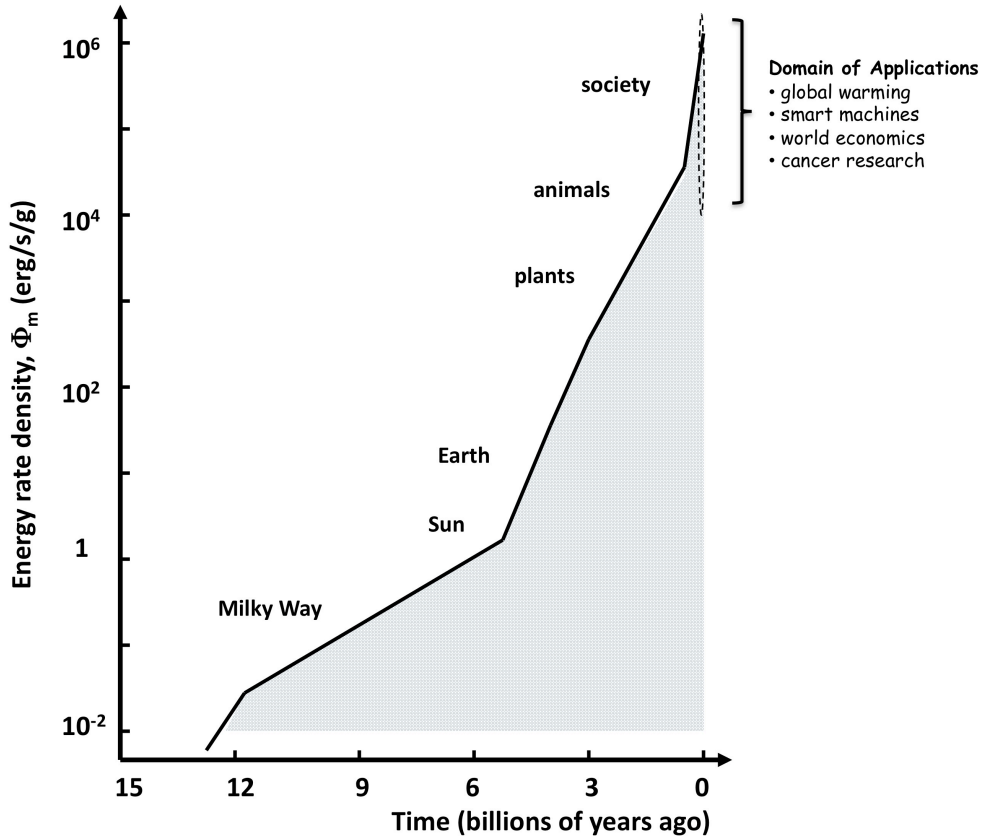

**Figure 1** — *Energy rate density, $\Phi_m$, for a wide spectrum of complex systems observed throughout Nature, displays a clear increase during ~14 billion years of cosmic history. The $\Phi_m$ values and their historical dates plotted here are estimates, all taken from and discussed in Paper I (Chaisson 2014). The thin dashed oval at upper right outlines the domain of $\Phi_m$ and time for the practical applications of cosmic evolution that are examined in the present analysis (Paper II).*

As a confirmed empiricist trained as an experimental physicist, I am skeptical of future forecasting because all such exercises entail much qualitative guesswork. Nor do I regard evolutionary events to be accurately predictable, even in principle, given that an element of chance always accompanies necessity in the process of natural selection; evolution is unceasing, uncaring, and unpredictable, all the while non-randomly eliminating over time the far majority of complex systems unable to adapt to changing environmental conditions (*cf.*, Paper I). Even so, it seems inevitable, indeed quite ordinary, that new forms of complexity are destined to emerge—some of them perhaps eventually supplanting humanity and its tools as the most complex systems known—just as surely as people took precedence over plants and reptiles, and in turn even earlier life on Earth complexified beyond that of galaxies, stars, and planets that made life possible. Here we examine not specific predictions, as much as four general trends that might affect humans in the near future: anthropogenic heat warming us, smart machines challenging us, world economics puzzling us, and medical disease afflicting us.

## II.  Climate Application of Cosmic Evolution

Today's civilization runs on energy for the simple reason that all ordered, complex systems need energy to survive and prosper. Whether stars, bugs, or civilization, it is energy that not only maintains the structural integrity of open, non-equilibrated systems but also keeps them functioning—helping them, at least locally and temporarily, to avoid a disordered state of high entropy ultimately demanded by the 2$^{nd}$ law of thermodynamics. Living or non-living, dynamical systems utilize flows of energy to endure. If stars do not convert gravitational energy into fusion, heat, and light, they collapse; if plants fail to photosynthesize sunlight, they shrivel and decay; if humans stop eating, we die. Likewise, society's fuel is energy: Resources come in and wastes flow out, all the while civilization goes about its daily business.

Closer analysis suggests a practical problem that inevitably arises for any energy-intensive society advancing during cultural evolution. This problem relates to the global warming that our planet now experiences—but not merely the familiar greenhouse-gas-induced warming that concerns us all. Even if humanity stops polluting our biosphere with greenhouse gases, we could still eventually be awash in too much heat—namely, the waste heat byproduct generated by any non-renewable energy source. Society is actually polluting Earth's air with heat, pure and simple, and although negligible now (<0.1°C) such waste heat is growing. Apart from the Sun's natural aging, which causes ~1% luminosity rise and thus ~1°C increase in Earth's surface temperature for each $10^8$ years (Sagan and Chyba 1997), well within a much shorter period of time our technological society could find itself up against a fundamental limit to growth. Thermodynamic modeling implies that within only a few hundred years, global waste heat could rise ~3°C—a temperature increase often considered a "tipping point" that could profoundly alter civilization as we know it, conceivably producing widespread drought, famine, and even mass extinctions (Chaisson 2008; Int. Panel Climate Change 2013). This biogeophysical effect has often been overlooked when estimating future planetary warming scenarios, and it is an example of how grand cosmic-evolutionary thinking can alert us to relatively near-term problems having potentially serious consequences for humankind. Fortunately, cosmic-evolutionary diagnostics can also help us avoid it.

**Rising Energy Use on Earth**
Of relevance to the much-debated issue of our planet's global warming is the often-ignored rise of energy usage among our hominid ancestors—a way of life that also characterizes today's digital society and will presumably continue well into the future. Energy rate densities can be estimated by analyzing society's use of energy by our relatively recent forebears, and the results illustrate how advancing peoples increasingly supplemented their energy budgets beyond the 2000-3000 kcal/day that each person consumes as food. Table 1 compiles values of $\Phi_m$ from Paper I (where these values were also plotted in Figure 8 of that paper) for several ancestral hominid and current human societies (*cf.*, Bennett 1976; Cook 1976; Simmons 1996; Christian 2003; Spier 2005, 2010; Chaisson 2008, 2011). Numerical values are rounded off; all are approximations, based on estimates available as of 2013.

A brief note on units: Much confusion results when different units are used to describe the energy budgets of various human groups and societies. Researchers from different specialties often use provincial (and sometimes non-metric) units to express the same quantity, and so Table 1 cross-correlates values of $\Phi_m$ in several sets of commonly used units: cgs metric units used in this (and its prequel) paper, SI units alternatively used by natural scientists, and per-capita values preferred by social scholars. All numerical values of $\Phi_m$ for each social system in Table 1 are closely equivalent.

**Table 1** — Societal Values of Energy Rate Density

| social system | emergence (ya) | $\Phi_m$ (erg/s/g) | $\Phi_m$ (kW/person) | $\Phi_m$ (kcal/day/person) |
|---|---|---|---|---|
| technologists in developed countries | 0 | ~$2 \times 10^6$ | 12.5 | 265,000 |
| modern citizens, on average | 0 | ~$5 \times 10^5$ | 2.6 | 55,000 |
| industrialists | ~200 | ~$3 \times 10^5$ | 1.6 | 35,000 |
| agriculturists | ~$10^4$ | ~$10^5$ | 0.6 | 12,000 |
| hunter-gatherers | ~$3 \times 10^5$ | ~$4 \times 10^4$ | 0.2 | 4,000 |
| australopithecines | ~$3 \times 10^6$ | ~$2 \times 10^4$ | 0.1 | 2,000 |

This table, and a much longer discussion of what it entails in Paper I, clarifies that per-capita daily energy usage in human history followed a slow and steady rise for long periods, then rising in a classic exponential growth more recently (*cf.*, Figure 8 of Paper I). All groups apparently needed a per-capita minimum of ~2000 kcal of food daily, which is likely an irreducible allotment for individual hominid survival. Hunter-gatherers used more energy to feed their animals, and agriculturists even more when conducting rudimentary trade among larger populations. In turn, industrialists required considerably more energy for production and transportation of goods. And today's technologists are yet more earnestly committed to energy use; virtually everything around us seems to run on energy.

Thus, energy *rates* have clearly increased over the course of recorded and pre-recorded history, but the cause of this rise is not population growth. These are per-capita (*i.e.*, per unit mass) rates of energy consumption resulting from the cultural evolution and technological advancement of our civilization (Energy Info. Admin. 2006). An underlying driver of much of this cultural advancement was not only greater total energy usage by society but also greater energy usage by each individual human being at each and every step of the evolutionary process. *In addition*, global population has grown and continues to grow, making clear humankind's formidable, ongoing, and rising energy demands, along with potentially grave consequences for environmental degradation and our future well-being.

The outcome is that energy rate density has continued rising right up to the present, as our modern world has become a humming, beeping, well-lit place—and there is no reason to think that it will stop anytime soon; increased per-capita energy use might well be a cultural imperative if the human species is to survive. Society's total energy budget will likely continue growing for three reasons: World population is projected to increase until at least late-21$^{st}$ century, when it might level off at ~9 billion people (U.N. Dept. Economic & Social Affairs 2008). Underdeveloped countries will mature economically, perhaps for more than a century until equity is achieved among the world's community of nations. And per-capita energy consumption (Table 1) will also probably continue rising for as long the human species culturally evolves, including that needed to air condition living spaces, relocate cities swamped by rising seas, and sequester increased greenhouse gases—all of which implies that even if the first two growths end, the third will indefinitely inflate society's total energy budget, however slowly.

**Heat By-products**

Current fears of energy shortfalls aside, our true energy predicament is this: we may eventually have too much energy in our Earthly environment. Unremitting and increasing use of energy from any resource and by any method necessarily dissipates as heat. Heat is an unavoidable by-product of the energy extracted from coal, oil, gas, atoms, and *any* other non-renewable source, including geothermal and nuclear. The renewable sources, especially solar, already heat Earth naturally, but additional solar energy, if collected in space and beamed to the surface, would also further heat our planet.

Regardless of the kind of indigenous energy utilized, Earth's surface is constantly subjected to heat generated by human society. We already experience a "heat-island effect" in big cities that are warmer than their suburbs and near nuclear reactors that warm their adjacent waterways. On smaller scales, everyday appliances produce heat owing to their thermodynamic inefficiencies: toasters, boilers, and lawn mowers all operate far from their theoretical efficiency limits. Electricity production is currently ~37% efficient, automobile engines ~25%, and ordinary incandescent lightbulbs only ~5%; the rest is immediately lost as heat. Even every Internet search creates heat at the web server, and each click of the keyboard generates heat in our laptops. Data processing of mere bits and bytes causes a miniscule rise in environmental temperature (owing to flip/flop logic gates that routinely discard bits of information). Individual computer chips, miniaturized yet arrayed in ever-higher densities and passing even higher energy flows, will someday be threatened by self-immolation!

Such widespread inefficiencies would seem to present major opportunities for improved energy conversion and storage, but there are limits to advancement. No device will ever be perfectly efficient, given friction, wear, and corrosion that inevitably create losses. Technological devices that are claimed to be 100% efficient are reversible and ideal, and they violate the laws of real-world thermodynamics; like perpetual motion machines, they do not exist. To give an example of a less-than-ideal gadget, today's photovoltaic cells currently achieve <20% efficiency, when optimized they might someday reach 40%, yet the absolute (quantum) limit for any conceivable solar device is ~70%. Overall in society today, about $2/3$ of all energy utilized is wasted and immediately dissipated into the environment.

Furthermore, it is not just waste heat per se (governed by the 2$^{nd}$ law of thermodynamics describing quality of energy) that is cause for alarm; according to the 1$^{st}$ law (energy conservation of quantity of energy), *all* energy used by our civilization (efficiently or not) eventually dissipates into the air at some temperature. That's why a better term for societal-induced heating is "anthropogenic heat flux"; society is heated not merely by inefficiently wasted heat as much as *all* of the energy used to sustain it. This does not allege that improvements in energy efficiency are unwelcome; in principle, higher efficiencies should cause less total energy usage and thus less heat and greenhouse gas pollution. However, in practice, advances in energy efficiency might backfire (Herring 2006; Gillingham 2013); as industry becomes more efficient, more goods are produced and consumed, often causing total energy usage to increase still more (or at least not decrease)—perhaps the best example being fuel-efficient cars, which cost less to run yet tend to get driven more while net energy savings often go unrealized (*cf.*, Sec IV below), and another is the increased use of medical tools and tests despite a constant stream of newly invented, more efficient devices. Experts now acknowledge that climate change is affected by economic growth twice as much as population growth [5]. Regarding today's civilization and its freakish economics, energy usage itself can have larger consequences than valued energy efficiency.

As we increasingly pollute Earth's air with heat, adverse climate change might conceivably occur *even in the absence of additional greenhouse gases*. How much energy can all of our cultural machines—automobiles, stoves, factories, electronics, etc.—produce before Earth's surface becomes hellishly uncomfortable? Thermodynamics offers an answer.

**Heating Scenarios**
The thermally balanced temperature T at Earth's surface is reached when energy acquired on the dayside of our planet equals that radiated away isotropically as a black body:

$$(k/r^2) \pi R^2 (1-A) = (\varepsilon \sigma T^4) 4\pi R^2.$$

Here, k is the solar constant at Earth (1370 W/m$^2$), r is the distance from the Sun (in A.U.), A is Earth's albedo (0.31), R is Earth's radius, $\varepsilon$ is the surface emissivity (0.61), and $\sigma$ is Stefan's constant. The result for Earth is 256 K, or -17°C, which is why we can be thankful for some natural greenhouse heating. That heating currently amounts to ~32°C, since the globally averaged T for Earth's surface is now measured to be 288 K (or ~15°C). This is the surface temperature value that has risen during the 20$^{th}$ century by ~ 0.7°C (Int. Panel Climate Change 2013).

Nature's power budget on Earth is dominated by the Sun. Compared to our planet's solar insolation of ~120,000 TW (absorbed by the land, sea and air, and accounting for Earth's albedo of 31%), our global civilization currently produces an imperceptible ~19 TW. But, with humanity's power usage on the rise (~2% annually; Int. Energy Agency 2008) as our species both numerically multiplies and culturally complexifies, society's energy demands by the close of the 21$^{st}$ century will likely exceed 100 TW, all of which will heat our environment.

Estimates of how much heat and how quickly it might rise rely, once again, on thermodynamics. Since solar flux scales as $\sigma T^4$, Earth's surface temperature will increase ~3°C when $(291/288)^4 = 1.04$, which means if only 4% more than the Sun's daily dose (~4800 TW) is additionally produced on Earth or delivered to Earth. Such estimates of energy usage sufficient to cause temperature increases are likely upper limits, hence the times needed to achieve them are probably lower limits, given natural greenhouse trapping and cloud feedbacks of the added heat. How far in the future, if ever, this might occur depends on assumptions (Chaisson 2007):
- If global non-renewable energy use continues increasing at its current rate of ~2% annually and all greenhouse gases are sequestered, then a 3°C rise will occur in ~8 doubling times, or ~280 years (or ~350 years for a 10°C rise).
- More realistically, if world population plateaus at 9 billion inhabitants by 2100, developed (Organization for Economic Cooperation and Development, or OECD) countries increase non-renewable energy use at 1% annually, and developing (non-OECD) countries do so at ~5% annually until east-west energy equity is achieved in mid-22$^{nd}$ century, after which they too continue generating more energy at 1% annually, then a 3°C rise will occur in ~320 years (or 10°C in ~450 years), even if $CO_2$ emissions end.
- If greenhouse gases continue soiling our atmosphere beyond the current ~410 ppm $CO_2$, all these projected times decrease.
- If only 4% *additional* solar energy that normally bypasses Earth is collected in space and beamed to the surface, its temperature would quickly rise 3°C (or 10°C for an additional 14% solar energy beamed here).

Even acceding that the above assumptions can only be approximate, the heating consequences of energy use by most means seem unavoidable within the next millennium—a period not overly long within a timeframe of real relevance to humankind on Earth—even if we were to end greenhouse-gas pollution and master nuclear energy. Changes in Earth's global albedo would not likely offset the added heating; even if all the world's glaciers (including Greenland) melted, their summed surface area is <1% that of our planet, and local albedo changes from dirty ice to typical landforms are not globally significant.

These estimates of global warming by waste heat have been generally confirmed in intricate models of Earth's atmosphere run on supercomputers (Flanner 2009). Although the total anthropogenic heat flux is currently negligible, statistically significant continental-scale surface warming of 0.4-0.9°C is forecast by the year 2100. Dissipated energy from urban heat islands is projected to increase from inner city centers to larger rural suburbs; climate simulations that neglect waste heat are deficient. Other recent computer modeling implies that thermal waste from 86 major cities accounting for nearly half of the world's energy consumption can disrupt atmospheric circulation, helping winds to warm other parts of the planet as well, and possibly providing an explanation for the heretofore anomalous winter warming (currently ~1°C) in the northern hemisphere during the last few decades (Zhang, Cai and Hu 2013).

Early musings about urban heat-islands date back decades (Budyko, 1969; Washington 1972), but their consequences are now more than theoretical. Such heating effects are among the best documented examples of anthropogenic change arising from increased urbanization today (*cf.*, Sec. IV); above-ambient heat has been detected in many large cities such as Tokyo, where its city streets are measured to be ~2°C warmer when air conditioning units not only suck hot air out of offices but also dissipate heat from the energy used to run such inefficient machines (Ohashi *et al.* 2007). Bangkok is another example of a big city whose discharged heat increases within its center where traffic is highly congested, causing only ~13% of the total energy input for transportation to be converted into useful work while the rest is released as heat (~3°C) into the environment (Moavenzadeh *et al.* 2002). London also experiences significant urban heating (up to 9°C on calm winter days in the city center) exacerbated by increased demand for electricity (Giridharan and Kolokotroni 2009). Waste heat generated by car engines, power plants, home furnaces, and other fuel-burning machinery already plays an unappreciated role in local and regional climates. Global climate effects, though still insignificant in the near future, seem destined to become relevant for Earth's atmosphere-ocean system within a century or two.

**Implications for Global Warming**

More than any other single quantity, energy has nurtured the changes that brought forth life, intelligence, and civilization. Energy also now sustains our society and supports our economy (*cf.*, Sec. IV), granting our species much health, wealth, and security. Yet the very same energy processes that have enhanced past growth also apparently limit future growth, thereby constraining solutions to global warming. Less conventional energy use, sometime in humanity's relatively near future, seems vital for our continued well-being, lest Earth simply overheat.

There is a way out of this dilemma—a resolution that allows continued, even rising, energy use without adverse heating. Thermodynamic waste derives mainly from non-renewable energy sources found on Earth. Whatever energy resource gets dug up from Earth's interior gets added to Earth's total thermal budget. That is, even if we embrace coal and sequester all of its carbon emissions, or employ nuclear methods (either fission or fusion) that emit no greenhouse gases, these energy sources would still spawn additional heat above what the Sun's rays create naturally at Earth's surface. By contrast, renewable energies, whose sole source is our Sun, are already accounted for in the thermal balance of our planet's air, land and sea, therefore their use would not additionally heat Earth's environment. Nor, incidentally, would energy derived from the solar-energy derivatives of wind, water, and waves. Furthermore, there is plenty of solar energy, far more than needed to power civilization today—as well as into the indefinite future. The ~120,000 TW of sunlight landing on Earth's surface each day equals nearly 10,000 times the power currently utilized by all humans and all of our machines combined; alternatively stated, Earth receives in only about one hour as much energy from the Sun as the human race currently uses in a full year.

Some colleagues claim that the $2^{nd}$-law degradation of our global environment on Earth presages the ultimate collapse of our technological society. In fervent contrast, I regard it as the single strongest scientific justification for adopting solar energy and its derivatives to power civilization going forward—allowing for significantly increased energy usage, including greater per-capita energy consumption, without additionally heating our planetary biosphere. We shall return to this topic—and this potential solution to one of humanity's foremost problems—in Sec. IV, while suggesting that solar energy can also best power the growth of our global economy perhaps indefinitely.

## III. Machine Application of Cosmic Evolution

Energy rate densities for human brains, society collectively, and our technological devices have now become numerically comparable in the early 21$^{st}$ century. If $\Phi_m$ is a genuine complexity metric, these are then among the most complex systems on Earth, indeed in the known Universe. As noted in Paper I (Chaisson 2014), I have no qualms about $\Phi_m$ values for some cultural inventions rising above those for human bodies and even brains; it is, after all, humans and their biological beings who build cultural systems, and so our creations that Nature never would have likely constructed without sentient beings might well function more complexly than our bodily selves. Accelerating cultural evolution is supported by a wealth of data, and rising complexity has now reached a crescendo with conscious beings, adroit machines, and their future intermingling (Figure 1). Yet the approaching, potential conflict between humans and machines is neither more nor less significant than many other, earlier evolutionary milestones as physical and biological systems changed and interacted along the arrow of time from big bang to humankind. The next evolutionary leap beyond sentient humans and their sophisticated tools will not likely be any more important (or troubling) than the past emergence of intricately complex systems. An oncoming cultural tipping point (or "singularity"; Kurzweil 2005; Eden *et al.* 2012) will cause increasingly smart machines to challenge humankind's dominant complexity as both speed and skill of computers rapidly accelerate—yet this clash between men and machines could conceivably create a positive symbiosis as each mutually benefits going forward.

Cultured humans and their invented machines are now in the process of transcending biology, a topic bound to be emotional as it rubs our human nerves and potentially dethrones our perceived cosmic primacy (Dick 1996; Dick and Lupisella 2009; Kelly 2010). The roots of this evolutionary milestone—perhaps it is a technological singularity—extend back at least to the onset of agriculture when our forebears began manipulating their local environs, and its effects are now quickly advancing as we alter both our planet environmentally and our being genetically. Even so, these changes—and their social outcomes—are probably nothing more than the natural way that cultural evolution proceeds beyond biological evolution, which in turn built upon physical evolution before that, each of these evolutionary phases being an integral part of the more inclusive cosmic-evolutionary scenario that also operates naturally, as it always has and likely always will, with the irreversible march of time.

**Humans Advancing and Machines Arising**

Rising energy expenditure per capita has been a hallmark in the origin, development, and evolution of humankind, an idea dating back decades (White 1959; Adams 1975). Culture itself is often defined as a quest to control greater energy stores (Smil 1994). Cultural evolution occurs, at least in part, when far-from-equilibrium societies dynamically stabilize their organizational posture by responding to changes in flows of energy through them. A quantitative treatment of culture need not be addressed any differently than for any other part of cosmic evolution. The result is that human societies typically utilize more energy per unit mass than biological organisms that originated before them, as explained in Paper I and compiled in Table 1 above.

As a benchmark against which to compare machines, consider the whole of modern civilization—namely, the totality of humanity going about its short-term social development as well as long-term cultural evolution. As noted in the previous section on climate change, ~7.2 billion people currently utilize ~19 TW to keep our complex 21$^{st}$-century society fueled and operating, so all of humankind together averages $\Phi_m \approx 5\times10^5$ erg/s/g. Table 1 further clarified the rise in $\Phi_m$ for our ancestors during the past ~10,000 human generations, displaying a steady increase in per-capita energy usage as our species culturally evolved from hunter-gatherers and agriculturists (~$10^5$ erg/s/g) many millennia ago to industrialists and technologists (~$10^6$) more recently. These many advances in energy usage have empowered human beings in countless ways by reducing drudgery, enhancing productivity, cooking food, generating light, providing transportation, powering industry, conditioning space for households and buildings, facilitating communications and operating computers, among untold technical tasks.

It is within more recent years that machines have culturally emerged, notably among them automobiles and aircraft that have become archetypical symbols of technological innovation worldwide. Paper I granted some perspective by documenting the rise in $\Phi_m$ (~$10^4$ to ~$10^5$ erg/s/g) as coal-fired engines of a century ago surpassed earlier steam engines of the Industrial Revolution, which in turn were bettered by gasoline-fired engines of modern times. For cars specifically, the value of $\Phi_m$ increased twofold during the past few decades, now averaging nearly $10^6$ erg/s/g. And aircraft that operate in three dimensions, and thus are more functionally complex than 2-dimensionally running automobiles, have $\Phi_m$ values that reach even higher—from the first airplanes (~$10^6$ erg/s/g) to today's commercial airliners (~$10^7$) to modern military jet aircraft (~$10^8$). More energy does seem to be required *per unit mass* to operate newer (even more efficient) vehicles, much as noted for the growth and complexification of so many other evolving systems in the Universe. This concomitant rise in $\Phi_m$ will almost certainly continue as machines fundamentally change their inner workings from heavy fuels to lightweight electrons and from mechanical linkages to small computers, thereby evolving degrees of upgrade yet unknown.

Another striking example of contemporary cultural evolution is the computer, which has perhaps replaced the automobile as today's premier technological icon. At the heart of every computer (as well as smart phones, digital cameras, ATMs, and many other consumer electronics) is the silicon chip whose complexity has grown geometrically in the past few decades, including stunning achievements in memory capacity and data processing speed. The number of transistors—miniature semiconductors acting as electrical amplifiers and logic gates—etched onto a single microprocessor has doubled every ~1.5 y, an advance obeying "Moore's law" (Moore 1965) marking each computer generation; Pentium-II chips of the 1990s that still power many home computers hold >$10^3$ times as many transistors (7.5 million) as the Intel-8080 chip (6000 transistors) that pioneered personal computers a (human) generation ago, and today's state-of-the-art chip, the Itanium-2, holds nearly 100 times still more. Chip development has been so rapid and its multiplication so pervasive that our post-industrial society may have already built more transistors than any other product in human history, including clay bricks.

Such stunning improvement in computer technology can be expressed in the same quantitative language expressed elsewhere in this analysis—here, the rate of energy flowing through computers made of densely compacted chips. In all cases, $\Phi_m$ values reveal, as for engines, automobiles, and aircraft above, not only cultural complexity but also evolutionary trends. (To make the analysis manageable, I examined only computers that I personally used in my career, except for the first and last device noted.) The ENIAC of the 1940s, a room-sized, 8.5-ton, 50-kW behemoth, transformed a decade later into the even larger and more powerful (125 kW) UNIVAC with ~5200 vacuum tubes within its 14.5-ton mainframe. By the 1970s, the fully transistorized Cray-1 supercomputer managed within each of its several (<1-ton, ~22 kW) cabinets less energy flow yet higher energy rate density as computers began shrinking. By 1990, desktop computers used less power but also amassed less bulk (~250 W and ~13 kg), causing $\Phi_m$ again to remain high. And now, MacBook laptops need only ~60 W to power a 2.2-kg chassis to virtually equal the computational capability and speed of early supercomputers. During this half-century span, $\Phi_m$ values of these cultural systems changed respectively: 6.4, 9.5, 32, 20, and 28, all times $10^4$ ergs/s/g. Although the power consumed per transistor decreased with the evolution of each newer, faster, and more efficient computer generation, the energy rate *density* increased because of progressive miniaturization—not only for the transistors themselves, but also for the microchips on which they reside and the computers that house them all. Currently, the world's most powerful supercomputer, the US Dept. of Energy's post-Jaguar Titan, devours 8.2 MW in its 200 cabinets (the weight of each classified but ≤1 ton), thus $\Phi_m \geq$ 5x$10^5$ erg/s/g.

The rise of $\Phi_m$ for computers generally parallels Moore's law and may be the underlying reason for it. Digital phones have continued this upward trend; the iPhone4 weighs ~130 g, charges at ~4 W rate, and typically uses ~1 GB (~3 kWh of electricity) for monthly wireless data transfers, making $\Phi_m \approx$ $3 \times 10^5$ erg/s/g—comparable to a $8-million *Cray* supercomputer of decades ago, yet now ~20,000 times cheaper and ~100 times faster. However, rapid, efficient computation does not always translate into energy savings; today's most advanced (metal-oxide semiconductor field-effect) transistors, with thin electronic gates only several atoms wide, actually consume more energy per unit mass (Wang *et al.* 2013), thus continuing the rise, albeit perhaps abated, of $\Phi_m$ in time. Some very fast, high-end cell phones have higher energy rate densities, in fact often use more *total* energy, than today's Energy-Star-rated refrigerators.

**The Human-Machine Interface**

Although these and other cultural $\Phi_m$ values often exceed biological ones, machines are not claimed here to be smarter than humans (despite the common terms "smart phones" or "smart machines"). Values of $\Phi_m$ for today's computers approximate those for human brains (*cf.*, Paper I) largely because they number-crunch much faster than do our neurological networks; even slim laptops now have central-processing units with immense computational capability and not surprisingly, in cultural terms, high $\Phi_m$ values. That doesn't make micro-electronic devices more intelligent than humans, but it does arguably make some of them more complex, given the extraordinary rate at which they functionally acquire and process data—and not least consume energy per unit mass. Accordingly, our most advanced aircraft have even higher $\Phi_m$ values than our most powerful computers. Modern flying machines rely on computers but also possess many additional, technologically novel features that together require even more energy density, in turn implying phenomenal complexity. That computers per se are amazingly complex machines, but not amazing enough for them to fly on their own, suggests that perhaps there is something significant—and inherently even more complex—about both living species and technical devices that operate in 3-dimensional environments on Earth; whether insects, birds, or jet aircraft, airborne systems exhibit higher values of $\Phi_m$ within their respective categories, probably more so to execute their awesome functions than to support their geometrical structures.

Much of this cultural advancement has been refined over many human generations, transmitted to succeeding offspring not by genetic inheritance but by use and disuse of acquired knowledge and skills. A mostly Lamarckian process whereby evolution of a transformational type proceeds via the passage of adopted traits, cultural evolution, like physical evolution, involves neither DNA chemistry nor genetic selection that characterize biological evolution. Culture enables animals to transmit modes of living and survival to their descendants by non-genetic, meme-like routes; communication passes behaviorally, from brain to brain and generation to generation, thereby causing cultural evolution to act so much faster than biological evolution (Dennett 1996; Blackmore 1999; Denning 2009). Even so, a kind of selection acts culturally, arguably guided by energy use (Chaisson 2011); the ability to start a fire or sow a plant, for example, would have been major selective advantages for those hominids who possessed them, as would sharpening tools or manipulating resources. The result is that selection accumulated newer technologies and systematically cast older ones into extinction, often benefiting humanity over the ages. It is this multitude of cultural advancements that has so dramatically escalated in recent times—advancements which, in turn with the scientific method that derives from them, enable us to explore, test, and better probe the scenario of cosmic evolution.

Figure 2 graphs many machine-related values of $\Phi_m$ computed above (and also in Paper I), as well as human-related values of $\Phi_m$ listed in Table 1; this graph derives from a more detailed analysis of the human-machine interface (Chaisson 2012) recently published among a collection of such papers (Eden, *et al.* 2012). Note that all these data pertain only to the uppermost part of the larger graph in Figure 1. Both modern society and our technical inventions are, in the cosmic scheme of things, extremely recent advances in the rising complexity of generally evolving systems in the Universe.

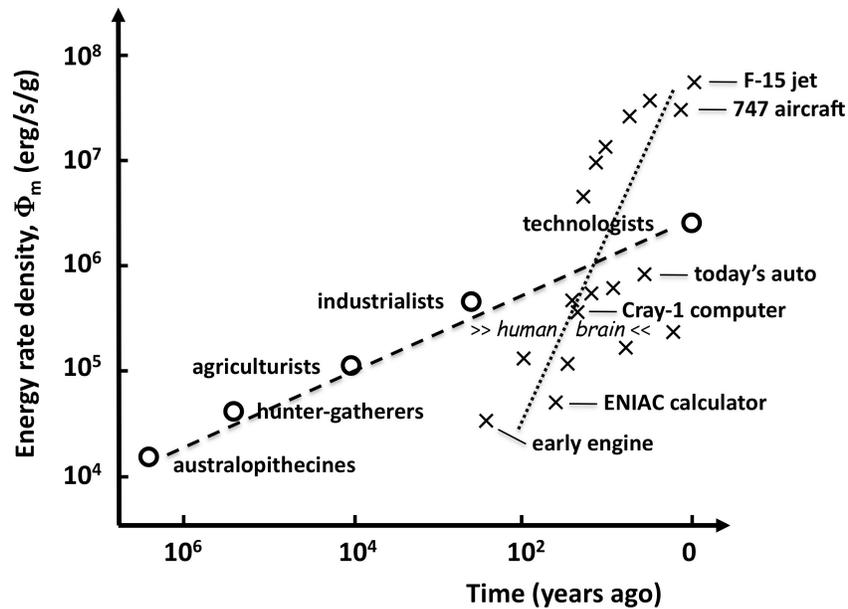

**Figure 2** — *Machines of the fast-paced 21st century not only evolve culturally, but also do so more quickly than humans evolve, either culturally or biologically. This graph shows some representative cultural systems that populate the upper part (within the thin oval) of the $\Phi_m$ curve of Figure 1. The time scale here covers only the past few million years, which is merely 0.02% of the total ~14 billion years of cosmic history. This is a log-log plot, allowing compact display of data computed in this paper for society (plotted as Os linked by a least-square-fitted dashed line) and for machines (Xs fitted by a dotted line) over millions and hundreds of years, respectively, in the same figure. The value of $\Phi_m$ for the human brain is also indicated—but note well that $\Phi_m$ is a proposed measure of complexity, not necessarily of intelligence.*

As noted in Paper I for many complex systems, $\Phi_m$ often rises exponentially only for limited periods of time, after which their sharp rise tapers off. Some but not all complex systems seem to slow their rate of growth while following a classic, sigmoidal (S-shaped) curve—much as microbes do in a petri dish while replicating unsustainably or as human population is expected to plateau later this century. That is, $\Phi_m$ values for a wide variety of physical, biological, and cultural systems grow slowly for long periods of time and then quickly for short durations, after which they level off throughout the shaded area of Figure 1 (whose drawn curve across all of evolutionary history is most likely a compound sum of multiple S-curves). Although caution is warranted to avoid over-interpreting the empirical data in Figure 2, such plateauing already seems evident for engines, aircraft, and perhaps society as a whole. This "maturing" of $\Phi_m$'s growth is discussed in greater detail elsewhere (Chaisson 2012; Modis 2012).

**Implications for Smart Machines**

Throughout the long and storied, yet meandering, path of cosmic evolution, many complex systems have come and gone. Most have been selected out of Nature by Nature itself—destroyed and gone extinct—probably and partly because they were unable to utilize optimal amounts of energy per unit time and per unit mass; in all aspects of evolution, there are few winners and mostly losers. Is humankind among the multitude of systems destined for extinction, owing perhaps to environmental degradation, societal collapse, or loss of control to machines?  Just what is the so-called technological singularity and can we objectively assess its implications in ways that go beyond subjective emotions?

Figure 2 allows a closer, numerical examination of the idea of a technological singularity—an occasion of some significance now perhaps underway during Earth's cultural evolution, which surely does transcend biological evolution.  Note that this graph is not temporally linear, rather fully logarithmic; as such, both (dashed and dotted) straight lines exhibit exponential growth—indicated individually for society advancing (plotted as Os, topped by modern technologists in developed countries) and for machines rising (plotted as Xs, topped by 3-dimensional, computer-controlled aircraft).  *Prima facie*, this plot does literally seem to display transcendence, as commonly defined "going beyond, surpassing, or cutting across," of machines over humankind; some machines already seem more complex (with higher $\Phi_m$) than the humans and their brains who created them.  This is often claimed to be an event beyond which human affairs cannot continue—akin to mathematical singularities beset by values that transcend finite limitations—one for which humankind and the human mind as we currently know them are ostensibly superseded and perhaps supplanted by strong, runaway, even transhuman artificial intelligence (von Neumann 1958; Kurzweil 2005).

The sum of the two curves in Figure 2 suggests faster-than-exponential growth for today's dominant cultural systems *en toto*—that is, the combined curve, dashed plus dotted (humans plus machines), sweeps upward on this log-log plot.  Cultural change is indeed rapidly accelerating and these $\Phi_m$ data prove it.  However, the data of Figure 2 imply no evidence for an event of singular import or uniqueness.  The technological singularity, which seems real and oncoming, may be central to beings on Earth (alas, especially threatening to our egos), yet it is only one of many notable events throughout natural history; this "singularity" is unlikely to be any more fundamental than many other profound evolutionary developments among complex systems over time immemorial.  The Universe has spawned many such grand evolutionary, even transcendent, events rightfully regarded as singularities all the way along the rising curve of $\Phi_m$ in Figure 1—including but by no means solely the birth of language (transcending symbolic signaling), the Cambrian explosion (land life transcending sea life), the onset of multicells (clusters transcending unicells), the emergence of life itself (life transcending matter), and even before that the origin and merger of stars and galaxies, among scores of prior and significant evolutionary events that aided the creation of humankind.

Men and machines need not compete, battle, or become mutually exclusive; they might well join into a symbiotically beneficial relationship as have other past complex systems, beyond which even-higher $\Phi_m$ systems they—and we—may already be ascending with change, namely, evolving a whole new complex state that once again emerges greater than the sum of its parts.  Conceivably, humankind could survive while becoming more machine-like, all the while machines become more human-like—these two extremely complex systems neither merging nor dominating, as much as coexisting.  After all, earlier evolutionary milestones that could easily have been considered transcendent singularities at the time—such as galaxies spawning complex stars, primitive life originating on hostile Earth, or plants and animals adapting for the benefit of each—did not result in dominance, but rather coexistence.  The wealth of empirical data presented in these two coupled papers suggest that singularities are part of the natural scheme of things—normal, frequent, yet broadly expected outcomes when concentrated energy flows give rise to increasingly complex systems in a perpetually evolving cosmos.

The technological singularity—one of many other singularities among a plethora of evolutionary milestones throughout natural history and highly unlikely the pinnacle or culmination of future cosmic evolution—fosters controversy because it potentially affects our human selves, even creating existential crises for those concerned about truly rapid change toward more technicality.  As some leaders now urge ethical constraints and regulatory restrictions on technological innovation and advancement, some people often wonder if we should strive to preserve our essential humanity and halt the growth of machines. Given the natural rise in an expanding Universe of the $\Phi_m$ curves in Figures 1 and 2, it would seem that we should not, indeed could not.

## IV. Economic Application of Cosmic Evolution

One of the hallmarks of cosmic evolution is that all complex systems are open, organized, and out of equilibrium. Nothing stable, fixed, or permanent pertains to them. Complex systems exist only temporarily, dependent largely upon energy flowing through them. Whether galaxies, stars, planets, life, society, or machines, all such increasingly complex systems utilize energy that grants them dynamically steady states of order and organization. If the energy acquired, stored, and expressed is optimum—neither maximized nor minimized, rather within different ranges for different systems of different masses (*cf.*, Paper I, Chaisson 2014)—then those systems can survive, prosper, and evolve; if it's not, they are non-randomly eliminated. In short, there is no such thing as a "balance of Nature," as ecologists formerly claimed for the biosphere on Earth. If Nature were actually equilibrated (thus its entropy maximized and energy minimized), stars, galaxies and life itself would not exist.

The economy, too, both local and global, is no different. The world economic system is just that—a *system*, in fact a very complex system with incoming energy and resources, outgoing products and wastes, and a distinctly non-equilibrium status. As for all complex systems, energy is likely key to the creation, growth, and operation, (as well as demise) of any economy; too much or too little energy utilization and the economy falters. Unfortunately, most of today's steady-state economists realize neither the essential role of non-equilibrium dynamics nor the importance of energy flows; most still apply decades-old equilibrium models that assume stability, balance, and input-output harmony in the marketplace. It is as though they prefer to regard the global economy as a closed system devoid of external forces, thus misrepresenting it as a relaxed, enduring combination of many internal parts. Economists' failure to recognize that local, regional, and global economies are driven far from equilibrium by robust energy flows is probably the principal reason why today's world economy is so unsettled.

### Non-equilibrated Global Economy

Economies are products of cultural evolution—social modes of organizing ecological space for greater yields and enhanced ends among humans having scarce means. Orthodox theories that regard the economic process as isolated and mechanistic (*e.g.,* Friedman 1953; Lancaster 1968; Samuelson and Nordhaus 2009)—even when revised to include dynamic effects (yet only for material flows; Leontief 1966) or thermodynamic insights (including inevitable soiling of environments; Georgescu-Roegen 1971)—still model the action of goods exchange as if economies were closed systems that are supply-demand equilibrated wherein rationally acting companies have perfect access to information, multinational networks are static, and the state of the economic system is computed using differential calculus. Yet much could be gained if economies were modeled as fully open systems that are optimized for product and wealth creation despite (in fact, largely owing to) their far-from-equilibrium status (Ayers 1994; Buchanan 2013; Arthur 2013). Such non-linear analyses aim to quantify the flows of energy needed directly and indirectly to provide durable goods and consumer services (Prigogine 1980; Odum 1996; Bakshi 2000). The bottom line—for this is economics, after all—does suggest that energy is the central currency of economies even more than money (or self-interest). Today's most successful businesses are all about speed of production (including design and manufacture) as well as turnaround of new and better products; high-tech communications and intense social networking help to accelerate ideas, research, and development. And nothing speeds things along more than energy, which is at the heart of all complex systems' evolution.

The novel interdisciplinary subject of ecological, or evolutionary, economics (Boulding 1978; Ruth 1993; Vermeij 2004; Costanza *et al.* 2009) embraces the core concept of energy flow (including material resources) under non-equilibrium conditions, just as other interdisciplines such as astrophysics and biochemistry (*cf.*, Paper I) have promoted energy as a principal organizing factor for many other complex systems. Understandably, social scholars concerned about natural scientists treading on their turf will likely resist notions of non-equilibrium, market gradients, and frequent institutional shifts, all of it implying economic life (and politics) on the ragged edge of chaos. Yet if we have learned anything from cosmic-evolutionary analyses, it is that all complex entities exist uneasily as though perched on an irregular arête, including pulsating stars, endangered species, and warring nations. It is, once again for the surviving systems among them, their dynamic steady-states that mix chance and necessity while wandering along the arrow of time toward greater complexity. That combination of randomness and determinism is also why realistic economies will never be predictable in detail, but will remain process-dependent, inherently dynamic, and always changing; all complex systems obey non-linear dynamics, precluding predictions far into the future. By contrast, economic equilibrium would signify a meltdown to nation-states and the financiers who seek to control them—a classic "heat death" of global markets and perhaps a collapse of technological civilization.

Neoclassical economists continue forecasting markets using linear methods, based on the premise that tomorrow's economy is a well-defined combination of features of today's economy. They view markets as inherently stable and self-regulating, often casting psychological risks and institutional factors in imposing mathematics typical of natural science, which economics is not (Geanakoplos 2008). By contrast, an emerging school of dynamic econometrics contends that commerce can be more accurately assessed when realizing that economies share common characteristics with all other complex systems in Nature—namely, all are disequilibrated systems forced out of balance by energy flows and environmental change, among other pressures. There is nothing self-regulating or self-organizing about economic markets; knowledge creation and product innovation are literally *driven* by energy. In particular, economies obey non-linear rules permitting rapid and unexpected fluctuations in the marketplace, much like abnormally violent storms can erupt in otherwise calm and ordinary atmospheric conditions; the difference between climate and weather affords perhaps a better analogy, the former providing long-term context for accumulated meteorological trends, the latter displaying short-term variations and occasional extremes in those trends. As with all complex systems, markets also commonly exhibit bifurcations—sudden changes in behavior of a system, some of whose small, natural variations amplify via positive feedback, much as in the rapid onset of fluid convection when input energy (heat) exceeds a certain threshold (Chaisson 2004). Such system behavior where more can become different (Anderson 1972), but is often just more within a complexity hierarchy, is usually orchestrated by flows of energy that do seem to cause, at least in part, some open, unstable systems to emerge as more complex entities (*cf.*, Paper I, §V). Furthermore, mathematical chaos can sometimes arise in systems, including economic systems (Motter and Campbell 2013), occasionally punctuating long periods of relative calm with brisk spikes of volatility (resembling, for instance, horses while rarely racing, galaxies briefly active, or microbes insatiably feasting). However, it would be a mistake to regard our market economy as confusingly chaotic, rather it dynamically evolves much as any complex system of many varied, interacting parts—although admittedly the former impression is often held by society during financial crises that have repeatedly disrupted human lives during the past few centuries.

We need to change our way of thinking about economics, much as it changed from Smith, to Mill, to Marx, to Keynes, to Friedman, all of who made, in turn, new and valuable contributions to the subject. Now is the time to take the next step forward in understanding the global economy by realistically modeling it, with empirical data, as a non-equilibrated, non-linear complex system that is rich in energy flows, continuously adapting, and subject to amplifying feedback—and no where are such avant-garde economic applications more pertinent than within and among our metropolitan areas.

**Cities as Economic Engines**

Cities (classified here as having >50,000 people and defined as "the form and symbol of an integrated social relationship . . . that concentrates culture and power in the community" [Mumford 1970]) are dynamic sources of innovation that enable socioeconomic development. Earth's worldwide economy is the sum of national economies, which, in turn, are networks of city economies. At even smaller, more local, levels, cities are diverse ecosystems of towns, companies, markets, and commercial enterprises whose operation and interaction mutually reinforce. Nation-states thrive economically when their urban social systems are vibrant, and that usually means robust energy flows; energy use and economic size are quantitatively correlated at all levels (although not everyone accepts that urbanization underpins world economic progress—*cf.*, Bloom *et al.* 2008). Cities are where most energy is utilized globally by civilization today because that is where most people live, now and increasingly so.

Urban systems are populous and dense, their structure and function organizationally intricate; almost everything about cities seems to be escalating—and complexifying. Cities are expanding and proliferating as humankind not only multiplies globally but also migrates from rural to urban areas. Although cities occupy <1% of Earth's land area, they now house ~55% of humankind and account for ~70% of all global energy usage; those latter percentages will likely increase to nearly 70% and 85%, respectively, within just a few decades as world population approaches 9 billion people (UN-Habitat 2006). In 1900, only ~13% of humanity lived in cities and hardly a dozen cities had more than a million residents (Modelski 2003); today >400 cities house this many people (mostly in Asia), and ~20 megacities have >10 million each, with Tokyo alone, for example, now having more residents than Canada and an annual economic output comparable to Australia. This flocking of people to cities at the rate of about a million new people per week is the greatest migration in human history and probably the most dominant cultural evolutionary trend of the 21$^{st}$ century.

Cities are as much a product of cosmic evolution as any star or life-form. As perhaps humanity's greatest social innovation to date, cities are culturally complex systems—"organic organized complexity" (Jacobs 1961)—that naturally emerge as people cluster for better health, wealth, and security (Jervis 1997). Historically, much of human progress has been closely linked to the emergence and development of cities; places like Uruk, Athens, Rome, Paris, among so many other famed locales, have often been at the forefront of social and intellectual advancement of humankind. Most established cities today are still evolving while hundreds of new ones are under construction, all of them trying (by means of cultural adaptation and Lamarckian selection) to achieve sustainable yet productive communities within Earth's human ecology (Odum 2007; Grimm *et al.* 2008). Much as for other complex systems, the makeup and operation—structure and function again—of cities (as well as of larger states and even bigger nations) can be analyzed in non-equilibrium, thermodynamic terms, for cities themselves are also energy-centered and dynamically stable (Dyke 1999). They acquire and consume resources, as well as produce and discard wastes, while providing many advantageous services: utilities, transportation, communications, construction, housing, medical, and entertainment, among a whole host of maintenance and infrastructure tasks. Although built culturally and not grown biologically, urban systems' principal activity can nonetheless be compared to metabolisms having energy budgets dependent on city size, location, culture, and history (Wolman 1965; Kennedy *et al.* 2007; Troy 2012).

Cities are surely voracious users of energy, not only to feed their many inhabitants but especially to provide the aforementioned amenities offered by city living. Of particular relevance to the present study, energy rate density values are high for individuals living in cities, $\Phi_m \approx 0.7(19\text{ TW})/0.55(7.2 \times 10^9 \text{ people}) \approx 3.4$ kW/person, or $\sim 7 \times 10^5$ erg/s/g on average for all cities of all nations; this agrees with estimates of the United Nations and World Health Organization that megacities typically use 300-1000 pentajoules per year to operate transportation, electrical, and climate control devices (Int. Energy Agency 2012). Nearly twice higher values of $\Phi_m$ pertain to some cities in developed countries (notably North America), a per capita power usage that residents of underdeveloped cities (currently with lower values of $\Phi_m$) will also likely achieve later this century. The above value of $\Phi_m$ exceeds by nearly 50% that for all humans generally (*cf.*, Table 1 in §II) since, as noted above, the heavily populated cities use more than their share of total global energy expended; alternatively stated, a whole city is greater than the sum of its many residents—yet another case of emergence among complexifying systems. Urbanization is a truly complex phenomenon since cities are highly heterogeneous, differing widely in population, buildings, and businesses (Brunner 2007); group size apparently does determine cultural complexity (Derex *et al.* 2013). Yet they all display a common trend: energy budgets for mature, developed cities are large, putting their $\Phi_m$ values near the top of the master plot of Nature's many varied complex systems in Figure 1. How humankind might continue meeting those high (and often growing) energy demands was discussed in §II; here we explore how such large urban energy flows impact economics both locally and globally.

Long-held assumptions and theoretical predictions have often maintained that larger, well organized cities foster greater efficiencies owing to shared infrastructure in high residential densities, implying that "economy of scale" saves energy (Glaeser and Kahn 2010; Puga 2010). However, contrary to such wishful thinking, recent data reported by several U.S. cities suggest that most urban systems are not so energy efficient—and the bigger they get, the more energy they proportionately need (Fragkias *et al.* 2013), totally and per capita. Such a diseconomic trend toward accelerated electrical consumption in bigger cities was earlier evident for a selection of German and Chinese cities (Bettencourt *et al.* 2007), but it was masked by distorted media reports that bigger cities always economize (they do for some shared utilities like cabling, plumbing, and roadways, but apparently not for each citizen's total energy needs). Thus, as cities double in population, they utilize *more* than twice the energy of their smaller selves. Not only does total energy usage increase with city size, but also per capita usage (thus $\Phi_m$) remains high and often even increases a little as well; individual residents of bigger cities use more energy than those living in smaller cities, and they use it at a rate proportional to or faster than cities' growth. Many other urban indicators also rise disproportionately faster with city size (including upsides like inventions, employment, wages, and social networking, but also downsides like crime, disease, noise, traffic, and pollution—*cf.*, Bettencourt *et al.* 2007; Bloom *et al.* 2008). All those urban benefits and detriments do cost energy—another inevitable result of thermodynamics' basic laws that not only help build systems but also degrade their environments; to be sure, cities are the largest producers of entropy on the planet. Per capita $CO_2$ emissions might decrease in our modern technological cities as automobiles travel shorter distances, but overall per capita energy use seemingly does not given increased electrical and other energies needed to run idling cars in congested traffic, air-conditioners to offset rising heat-island effects noted in §II, and battery-chargers for a wide array of smart machines noted in §III—just glance at today's electricity, phone, or cable bills. The probable reason for these urban energy supplements in the bigger cities—implying greater, not less, complexity as cities grow—is their enhanced networking (among many other valued urban qualities), which in turn fosters increased numbers and diversity of interactions within cities' burgeoning populations. Such social engagements are welcome and beneficial, and along with the underlying influence of energy use (both in absolute terms and on a per capita basis) strongly aid knowledge creation and product innovation for the human species. This is cultural evolution at work in its most rapid and vigorous way to date, yet fundamentally no different than for other aspects of cosmic evolution; humans cluster into cities much like matter clusters into galaxies, stars, and planets, or life itself into bodies, brains, and society; all these complex systems are basically governed by the same general principles of thermodynamics that guide energy flows, as quantitatively delineated by rising $\Phi_m$ in Figure 1.

Complete, current, and accurate energy data for individual cities are very hard to find; urban managers keep few records of this neglected diagnostic, which is usually compiled for states and nations (Decker *et al.* 2000; Int. Energy Agency 2008). Figure 3 plots $\Phi_m$ in two ways, spatially and temporally: in part (a) $\Phi_m$ is shown rising with size of cities generally (dashed line, adapted from Bettencourt *et al.* 2007, such that $\Phi_m \approx \Phi_{m0} P^{0.1}$ where P is population); in (b) $\Phi_m$ is shown rising (or leveling off) with time for the specific cities of Sidney and Toronto in two of the most energetically expensive continents, and for Hong Kong within a developing country (Kennedy *et al.* 2007). The dashed, upward trends of these graphs agree with the hypothesis that complexification of virtually all organized systems display increasing $\Phi_m$. Individual stars, for example, increase their $\Phi_m$ values while evolving and complexifying (*cf.*, Paper I), much as the data imply for developing cities in Figure 3b; and multiple stars of different sizes also show that $\Phi_m$ increases with mass (Chaisson and McMillan 2014), akin to the case of cities in Figure 3a. However, numerically, $\Phi_m$ for cities exceeds that for stars by many orders of magnitude, in keeping with the intuitive impression that cities are much more complex than stars. Energy rate density holds as a general measure of system complexity, just as it has for so many other complex systems that have emerged throughout cosmic history, from big bang to humankind. As cities culturally evolve, they become more massive, dense, and complex. The pace of big-city life feels more energetic because it is. Earth's cities, too, are an integral part of Nature.

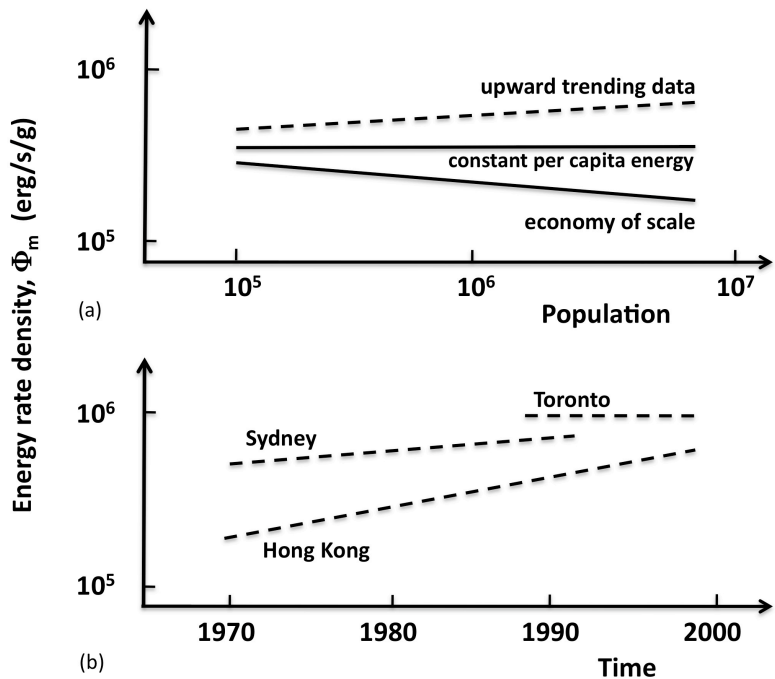

**Figure 3** – *Bigger cities generally complexify as they evolve, and that implies, as for all complex systems, increased energy rate density. (a) Sparse data for real cities (dashed line for electrical energy used in some Chinese, German, and U.S. cities) suggest that $\Phi_m$ neither stays constant (level scaling of per capita energy use with population size) nor decreases for economy of scale; rather, $\Phi_m$ continues rising as cities grow, if only slightly. (b) Energy expenditures of urban residents generally rise as their cities grow (Sydney somewhat, Hong Kong more so), at least until reaching maturity when their city values of $\Phi_m$ level off (as for Toronto, which might have already plateaued along an S-shaped growth curve). Cities are energy hungry—and the bigger they get the more energy their residents use, both totally and per capita.*

As cities, towns, and metropolitan areas bioculturally evolve, each must change and adapt, often rapidly so—in built infrastructure, consumer lifestyle, and human behavior. Surprisingly, cities' viability may not depend upon improved efficiencies; frequent assertions that energy efficiency confers competitive advantages (Glaeser 2011; Troy 2012) are dubious. Rather, cities can perhaps best thrive economically in the 21st century when its many city dwellers take full advantage of increased energy availability (as well as solve inevitably increased environmental degradation). The laws of thermodynamics demand adherence: Cities able to manage their energy budgets optimally are most sustainable in the long run and will likely survive; other cities using too much or too little energy—beyond an optimal range of $\Phi_m$, as for all complex systems—will likely be naturally selected to terminate. A fine line separates existence from extinction all through Nature—and like most Phanerozoic species that became extinct, most new companies underlying urban economies fail; each year >10% of all U.S. companies disappear.

Despite the apparent lack of energy savings (even per capita, see Fig. 3) as most cities grow and develop, opportunities abound to improve cities' energy efficiency, thus slowing the rise of $\Phi_m$—and perhaps halting it altogether as cities "mature." That might be why, for the entire U.S. nation where ~84% of its citizens already live in metropolitan areas (up from ~28% a century ago, as judged by the U.S. Census Bureau; *cf.*, Wilson *et al.* 2012), the rise of $\Phi_m$ has slowed in recent years (Energy Info. Admin. 2011). Economically, energy efficiencies are welcome because energy usually costs less when we use less. Innovative ways to design and implement energy-efficient city projects are offered by many leading organizations, including the World Bank (2008), the United Nations (UN-Habitat 2006), and the International Energy Agency (Jollands *et al.* 2008). A reasonable expectation for cities generally is that the rising temporal dependence of $\Phi_m$ will eventually turn over in a sigmoidal S-curve, much as noted following Figure 2 for any complex system's origin, growth, and maturity. This might have already occurred for some energy-intense North American cities, such as Toronto (see Fig. 3b). However, $\Phi_m$ values for sustainable cities will not necessarily, and perhaps not likely ever, decrease—a common misconception among urban analysts—thus the need for yet more energy to operate our cities, our society, indeed all of civilization for as long as these complex social systems endure.

Failing U.S. cities like Detroit, Buffalo, and New Orleans, among several others worldwide, such as Juarez, Pyongyang, and Mogadishu, are not immune to these statements. For example, with little industry, huge debt, social mismanagement, unemployment >20%, and a ~30% decrease in population in the past decade, Detroit is a naturally collapsing city on the brink of operational ruin; New Orleans might also be doomed eventually, as for so many cities built on the banks of waterways. Without government intervention (mostly as monetary handouts for energy-centered tasks), cities with decreasing $\Phi_m$ values will likely end; they will be culturally selected out of the category of urban entities (or at best urban-renewed as smaller, less complex social systems). Great cities have indeed fallen, including for example Sumer's Uruk and Ur, Egypt's Memphis and Mohenjo-Daro, Persia's Babylon, ancient Rome, Troy, Angkor, Teotihuacan, among many others that came and went throughout recorded history. Some vanished via conquest, disease, or environmental disruption, likely unable to manage optimally their energy budgets (wars utilize too much energy, famine too little, and often not even economic revival prevents collapse—*cf.*, Tainter 1988; Diamond 2004). Energy-based anthropological analysis of Mayan Indian society draws a distinction between vertical (upward rise of $\Phi_m$) and horizontal (leveling of $\Phi_m$) evolutionary strategies, implying that sometimes minimal (or zero) complexification is favored provided the social system doesn't devolve into collapse itself (Adams 2010). Failure is a frequent outcome in the natural scheme of cosmic evolution for all complex systems, and urbanism is no exception. It is too early to know if cities will survive as ordered phenomena, quite impossible to predict specifically where the curves of Figure 3 are headed. Cities are among the youngest advances of cultural evolution, thus particularly susceptible to physical, biological, and social constraints that could fundamentally change, or even eliminate (via selection), those very same cities.

I conclude that as cities evolve, some infrastructure efficiencies are realized but energy savings may not be one of them. Total energy utilized rises for each growing city and so does per capita energy usage for many urban citizens; generally the larger the city, the more hungry it is in nearly every energy sector (transportation perhaps excepted). To survive, cities of the future will not necessarily need to become more efficient; rather, they will need to acquire more energy—not only more total energy for their urban economies but also more per capita energy for virtually each and every resident. As noted at the end of §II, only the Sun and its associated renewable sources of wind, water, and waves can possibly provide humanity with the needed energy.

**Implications for World Economics**

From a cosmic evolutionist's viewpoint, the global economy is all about energy. Success and sustainability of cities, which comprise the core of our planet's economy, are closely and mostly tied to energy. Although energy use is now high in the cities where most people live, even more of it will likely be needed not only to lift developing nations out of poverty but also to increase the standard of living for everyone. Economics, in particular, will need to direct greater flows of energy into, within, and among cities. Energy budgets are destined to rise in all urban areas, including per capita energy usage; rising $\Phi_m$ seems a cultural imperative and pragmatic economic behavior needs to learn to manage it.

Economists and their "dismal science" are easy targets for criticism today, especially given their historic inability to effectively manage (or even explain) world markets (Marglin 2008). While most mainstream practitioners argue that technological innovation can ensure unchecked growth as the best way forward (Barro and Sala-i-Martin 2003; Mankiw 2010), opponents counter with Malthusian pessimism that societal growth is inherently restricted owing to resource shortages (Malthus 1798; Brown *et al.* 2011). I was formerly among the latter, regarding energy supplies as insufficient to power civilization robustly and limitlessly. Yet, as suggested in §II of this paper, cosmic-evolutionary analysis urges recognition that humanity's future depends largely on adoption of solar energy, plenty of which is available to power economic growth practically forever. This is not merely the stance of an astrophysicist looking to the sky for solutions to earthly problems, nor is it an endorsement of economists who advise better efficiencies when they should be recommending optimal energy flows for humanity's future well-being. Use of solar energy seems the only seriously viable way that our technological society can avoid collapse, continue evolving, and ultimately be selected by Nature to endure indefinitely.

With the Sun as society's main energy source, demand for it will soar but energy prices should not since its supply is plentiful. Our future economy can be built on solar energy without the boom and bust of a recessionary economy that most economists nowadays try to keep fixed and equilibrated. No longer would energy issues arise only when market crises impact society; with a solar economy, energy will be so abundant and (eventually) cheap that it will safely guide civilization independently of erratic decision-making. Such a global solar-based economy would still produce numerous goods and services, but the one resource—energy—that underlies all complex systems, including human society, would no longer be subject to geopolitics, revolutions, or greed. And urban energy metabolism can become an earthly virtue, shepherding the structures and functions of our cities and their residents without further degrading surrounding environments. The Sun can grant us the freedom to use much more energy, all the while freeing us from reliance on dirty fossil fuels. That is the economy—an open, non-equilibrated, solar economy—to which we should aspire going forward.

Humankind needs to raise its energy acumen—to think big and adapt broadly to what fundamentally drives human society. That driver is not likely information, the internet, money, or any other subjective label that theorists and pundits often preach; rather, all complex systems, including society, are root-based on energy, and objectively so as suggested in these two papers. Some researchers do recognize that urban energy metabolism is an economic issue and not merely an environmental one, but their premise, much in accord with currently fashionable equilibrium economics, urges cities to become more efficient by conserving energy (Troy 2012). This seems unrealistic in today's energy-centered society, which even with efficiency gains might require yet more energy, implying that energy savings could ironically translate into higher consumption; "Jevons' paradox" implies that as efficiency rises for any device, market pressures tend to lower its price, thus increasing demand for it—which is why many people who buy cars with good mileage ratings often drive more and those who are comfortable with smart gadgets tend to own more of them, ultimately often using just as much and sometimes more total energy, and also why 21$^{st}$-century citizens use vastly more per-capita energy than our ancient forebears even though modern machines need only a fraction of antiquity's "horsepower" for any given task (Jevons and Flux 1965; Herring 2006). Cosmic evolution and its undeniably upward trends near the tops of Figures 1-3 advise copious amounts of additional energy to flow within society (especially if that energy derives from the Sun, is relatively cheap, and minimally degrades Earth's environment), thereby allowing disequilibrated global economics to flourish, expand, and further complexify society with city life as its heart (*cf.*, Cleveland 1984; Batty 2008, 2013).

Quantitative correlations between energy use and economic development sometimes elicit the query, which caused which? The answer seems clear: Just as energy clearly drives metabolisms in organisms and energy flow is key to the thermodynamic rise of complexity among all ordered systems (*cf.*, Paper I), energy surely affects economic growth within and among nations. On smaller scales, that energy is the principal source of many local economies is evident by examining the gross domestic product (GDP, a standard measure of prosperity) of several U.S. states. The overall GDP for the entire U.S. grew in 2012 by 2.5%, much of that caused by the fastest-growing states that have embraced the energy industry. For example, Texas enjoyed 4.8% economic growth mainly owing to its energy production, as did West Virginia (3.3%), and North Dakota (13.4%), the last of these leading all states with its newly booming mining, drilling, and fracking operations; by contrast, other states that are not major players in the energy business, such as Connecticut, Delaware, and New Mexico, reported low (<0.2%) GDP growth rates (Bur. Econ. Anal. 2013). On larger, national scales, unambiguous correlations between earned income and energy use further implies strong connections between per capita energy consumption and GDP (Brown *et al.* 2011). Manufacturing, trade, finance, and the insurance industry all contributed to rising national GDP and recovery from the Great Recession of 2008-11, but none greater than the energy sector. This hardly surprises given that energy usage undergirds modern society like no other factor and will probably continue doing so for as long as society survives. There is nothing more fundamental, nor essential, for civilization's viability than energy, and the need to keep its energy rate density relatively high and optimized.

None of this analysis—or even a complete articulation of non-equilibrium economics, which no one has yet achieved—claims that economics is predictable or that markets are controllable. General trends can be identified (as in the many graphs of these two papers), yet specific predictions resembling the precisely deterministic Newtonian trajectory of a bullet are impossible. As with all other aspects of cosmic evolution, local and global economies depend on both chance and necessity as non-equilibrium thermodynamics goes about its business of guiding energy flows through complex systems—which for the case of human society is ourselves mostly within the vibrant and expansive cities of planet Earth. City policy-makers would be wise to welcome integrated, evolutionary review of societal and environmental challenges now confronting human settlements. Urban planning and climate mitigation should include the realistic likelihood that people in cities will use large quantities of energy, indeed increasing amounts and rates of energy, for the foreseeable future.

## V. Medical Application of Cosmic Evolution

Cancer is a systems disease, and recent efforts in systems biology have seen a resurgence of studies of the metabolisms of abnormal cells and aberrant tumors. Cancer research today is no longer guided solely by the general assumption that tumor cells' behavior depends on DNA sequences and a reductionist, genomic-based focus on tumors housing bad cells. Rather, tumor-cell conduct is increasingly viewed holistically in ways that seek to diagnose the functioning of whole systems within their extended micro-environments—and energy-centric metabolic mechanisms are at the core of this recent reevaluation (Levine and Puzio-Kuter 2010; Dang 2012; Lazar and Birnbaum 2012; Seyfried 2012). Today, more physicists are collaborating with biologists (and physicians) in the "war on cancer" (Gravitz 2012; Davies 2013).

Overall, invasive tumors metabolize much like normal cellular systems, interacting with their neighboring surroundings while acquiring energy, producing mass, and secreting wastes. Key differences are that cancer cells utilize greater energy and usually operate outside the optimal range of energy rate density for their host organisms, often growing and proliferating by sending their metabolic cycle into overdrive. It is now accepted that dysregulated metabolic change in cancer cells is a key promoter of tumor formation (Tasselli and Chua 2012; Schulze and Harris 2012).

Despite some notable advances in medical oncology (Hanahan and Weinberg 2000, 2011), mortality rates for major cancers have not improved much during the past few decades (Marshall 2011). Unorthodox strategies toward understanding (and treating) cancer are now desirable, especially if they involve metabolic intervention. Given cosmic evolution's central premise that complex systems can be consistently and uniformly characterized in terms of their energy flows per unit mass, it is not inconceivable that this natural-science interdiscipline might aid in identifying new ways to address one of modern medicine's foremost challenges today—the search for cancer's cure.

**Cell Thermodynamics**

Physicists often regard cells as physico-chemical systems wherein the whole anatomy is vital and energy is central, in contrast to the more reductionist approach of molecular biologists, with their DNA-centered viewpoints and huge genome databases. Systems biologists seem to operate somewhere in between. Energy is an attractive concept mainly because it is well understood, unambiguously defined, and directly measurable. By contrast, entropy as a quantitative diagnostic of complex systems is not as useful mainly because, unlike energy and entropy's oppositely trending energy rate density, empirical measures of entropy are virtually impossible. Nonetheless, theoretical checks that all such systems obey the fundamental laws of thermodynamics are germane and here the salient calculations are made for animal cells metabolizing.

Cells are organized, non-equilibrium systems, open to energy flows—isothermal systems lacking temperature gradients and therefore unable to be powered by heat alone. Unlike a battery or some other physical system that converts chemical energy to thermal energy and thence mechanical work, biological systems convert chemical energy directly into mechanical energy used to run metabolic functions such as digesting foods, synthesizing biochemicals, and contracting muscles. Even so, an intermediate step is required to power life—whether bacteria, plants or animals—suggesting (owing to its commonality) that this advance must have evolved early on in the history of life. As living systems metabolize incoming sugars and other carbohydrates, they produce adenosine triphosphate ($C_{10}H_{12}N_5O_4[PO_2OH]_3H$, or ATP for short), the fuel-like chemical carrier of energy from the site in a body where food is consumed to the site where it is used. It is within this molecule where energy once acquired is then stored, ultimately to be expressed (by releasing chemical bonds rich in energy) when organisms do work (Lehninger *et al.*, 1993).

ATP is the primary agent that powers work among all forms of life on Earth. When food is eaten by living systems, animal or vegetable proteins are broken down by digestive enzymes into their constituent amino acids. This is an entropy-increasing process because somewhat ordered, larger molecules are converted into many smaller ones having more randomized spatial arrangements and increased freedom of motion. Protein is then synthesized by combining those acids in the correct order and type to be useful to the system. This, in turn, is an entropy-decreasing process, which would not occur without the entropy-increasing combustion of fuel as noted above.

The balance sheet for part of the metabolism running the chemical engine of a living organism can be tallied in the following way: Consider a single reaction that is representative of ATP participation in cellular metabolism, namely the assembly of glycogen, a long chain of ring-shaped glucose ($C_6H_{12}O_6$) molecules formed by photosynthesis and linked as a polymer macromolecule; this is specifically where animals store their carbohydrate (hence energy) supplies, most of it concentrated in the liver and muscles. Chemical analysis (Stryer 1988) shows that the simplest case of linking only two glucose molecules yields a decrease in entropy, S, for this is a process that builds up order within the system:

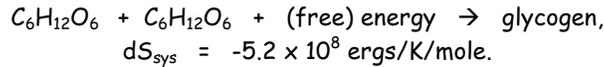

$$C_6H_{12}O_6 \; + \; C_6H_{12}O_6 \; + \; \text{(free) energy} \; \rightarrow \; \text{glycogen,}$$
$$dS_{sys} \; = \; -5.2 \times 10^8 \; \text{ergs/K/mole.}$$

The free energy ($+3.8 \times 10^{-13}$ erg, or $+0.24$ eV), which is "freely" available in the thermodynamic sense to do work, that drives this endergonic synthesis comes from the exergonic conversion of an ATP molecule to ADP for each glucose added to glycogen; reacting spontaneously with a single $H_2O$ molecule ("hydrolysis"), ATP's energy-rich bond is broken with its terminal phosphate group ($PO_3H$), thereby forming a more stable system that releases the energy ($-4.8 \times 10^{-13}$ erg, or $-0.30$ eV) needed to power the above vital biochemical reaction within an organism. And here, computations show that entropy of the surrounding micro-environment increases, for this is a process that reduces order:

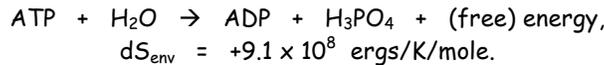

$$ATP \; + \; H_2O \; \rightarrow \; ADP \; + \; H_3PO_4 \; + \; \text{(free) energy,}$$
$$dS_{env} \; = \; +9.1 \times 10^8 \; \text{ergs/K/mole.}$$

In this way, ATP acts as an intermediate fuel during respiration, helping to make large molecules from smaller ones, indeed to convert simple molecules into more complex cell constituents. The two-step process outlined here is a simplified version of the intricate reaction-catalyzed pathways involving ATP, yet the initial reactants and final products as well as the numerical entropy gain for the complete transaction are exactly the same. Biosynthesis of this sort resembles the formation and folding of a protein by the clustering of amino acids or the assembly of a ribosome, both of whose increased organization is more than offset by the decrease in the order of the surrounding water molecules. Not surprisingly, the net effect (system + environment) of all these coupled biochemical reactions is an entropy increase, as required by the 2$^{nd}$ law of thermodynamics.

Metabolic rate is actually the rate of ATP production. The basal rate is the minimum rate at which an organism uses energy to maintain structural integrity—to stay alive. Active (or field) metabolic rates are greater and allow organisms to function. Although lacking thermal gradients, living systems do have non-equilibrium concentration gradients across membranes—and it is this characteristic of life that requires a constant influx of energy and a consequent production of entropy. For normal, healthy cells, that entropy gets dumped into the micro-environment beyond the cells and eventually the larger environment beyond the body housing the cells; for cancerous cells (see below), entropy degrades cells per se as well as potentially everything around them. During carcinogenesis, cancer cells do not disobey the 2$^{nd}$ law, or even circumvent it. Rather, they fully obey this celebrated, inviolable principle of Nature, unfortunately increasing entropy in the affected cells as well as their surrounding tissues, thereby often taking a serious toll on both the structure and function of a host organism.

### Energy Rate Density for Human Cells

Energy rate density, $\Phi_m$, for adult humans of 70-kg mass going about their normal routine while consuming ~2800 kcal each day (~130 W) equals ~$2 \times 10^4$ erg/s/g. This is a mean mass-specific metabolic rate within a range of values for humans, as for all complex systems. The basal rate for a person at rest is ~40% less whereas the active rate during exertion such as running or swimming is several factors more. Given that there are ~$10^{14}$ cells in the human body, some colleagues then reason that each individual cell has a hundred-trillion times smaller $\Phi_m$, or ~$10^{-10}$ erg/s/g. But such a divisional analysis is incorrect since $\Phi_m$ is a density quantity. In fact, each cell in the human body has a value of $\Phi_m$ comparable to that of the whole human body—much as a rock with a uniform density of 5 g/cm$^3$, if broken into many pieces, guarantees that each smaller piece retains a density of 5 g/cm$^3$.

This human cellular estimate of $\Phi_m$ can be confirmed more directly. Neglecting the microbial cells in our body (but see next section), since, despite outnumbering our mammalian cells by ~10:1, the mass of each microbe approximates 10$^{-12}$ g and thus altogether amounts to less than a few percent of our total bodily mass, we then find: When our daily consumption of ~2800 kcal is utilized by ~10$^{13}$ cells, each one uses, on average, ~10$^{-11}$ W and since mammalian cells average ~10$^{-8}$ g, then $\Phi_m \approx 10^4$ erg/s/g. Such high values of $\Phi_m$ for a single cell—in fact, thousands of times greater than for the Sun—should not surprise us since the transport of fluid across cell membranes requires much energy per unit mass.

More specifically, of the ~10$^{13}$ mammalian cells in the human body, $\Phi_m$ varies somewhat among our ~200 different types of cells, depending on the bodily part examined. Gut, bone, and muscle organs have a few factors <10$^4$ erg/s/g, whereas our brain displays roughly an order of magnitude higher, ~1.5x10$^5$ erg/s/g. Much like our host bodies in which they reside, individual cells vary greatly in size, scale, mass, type, and function; variation is normal in biology and essential in evolution. For example, cells of the human oocyte (in the ovary), macrophages (in blood and tissue), and adipocytes (in the abdomen) often grow—that is, physically build mass by adding DNA, proteins or lipids, and not merely dividing—to several times their typical 10-20 $\mu$m diameters; they thus have more mass (which scales as the cube of their crossection) and less $\Phi_m$. Additional cell growth and proliferation in animals is especially evident for organisms that are diseased, as noted below, implying even lower $\Phi_m$ (thus higher entropy) than for cells enjoying normal, healthy physiology.

**Microbial Metabolism**

Since unicells became multicells and thence more complex organisms as life ascended with modification during biological evolution, the earlier rise of single cells is often associated with the domain of chemical evolution between organized systems that are living and simpler ones that are not; this implies, based on the curve of rising $\Phi_m$ in Figure 1, cellular values of $\Phi_m$ in the range of tens to hundreds of erg/s/g. Experiments for respiring microbes often report higher energy rate densities of order 10$^6$ erg/s/g, but such measures are difficult to gauge since microbes differ dramatically among active, normal, and dormant states. Consider the well-known unicell, *E. coli*, a 2-$\mu$m-diameter bacterium populating the human intestine, each with a mass of ~2x10$^{-12}$ g. At peak activity in ideal laboratory cultures, *E. coli* utilizes energy maximally, reproducing every 22 minutes—and if left unchecked, with sufficient resources, would yield in a single day a progeny of ~10$^{28}$ g, which is roughly the mass of the entire Earth! That obviously doesn't happen, not even close. In reality *E. coli* hardly ever consumes energy uncontrollably at maximum rates under ideal, *in vitro* conditions, in fact much more often metabolizes at normal or even basal rates, implying that $\Phi_m \ll 10^6$ erg/g/s; when time-averaged *in vivo*, *E. coli*'s $\Phi_m$ value usually falls between hundreds to thousands of erg/s/g. This seems anecdotally consistent with an acclaimed observation that normally growing bacteria produce very little heat (Monod 1971), yet biologists with short-term grants seemingly lack patience to examine ordinary microbes that slowly consume energy, much as astronomers find active galaxies boring during their more common, inactive phases while stingily feeding their central black holes.

Wide differences between active and basal rates, as well as field and laboratory studies of many related living phenomena, abound in the literature. Hibernating animals are good examples, such as black bears that feed insatiably during fall while gorging ~20,000 kcal of berries for ~20 hours each day, spiking their $\Phi_m$ an order of magnitude to >10$^5$ erg/s/g, after which that value plummets during hibernation; freshwater turtles are another species that exhibit much higher $\Phi_m$ values when functioning during warm weather than while barely surviving in winter at the bottom of frozen lakes devoid of any $O_2$ (Madsen *et al*. 2013). Likewise in a related context, cheetahs' speeds often exceed 100 km/hour while in captivity where they are unaccustomed to hunting, but in the wild they tend to run at more moderate ~55 km/hour, energizing to higher speeds only at the final moment of the kill (Wilson *et al.* 2013). Overly active states are not the norm, rather the exception, throughout much of Nature.

The common soil bacterium, *Azotobacter*, is another voracious heterotroph when resources are abundant; its high $O_2$-consumption implies ~$10^6$ erg/s/g, but like *E. coli* these extraordinary bacteria do not always, or even often, maximally respire, making their time-average values of $\Phi_m$ much smaller and more comparable to many other aerobic soil bacteria, >80% of which are nearly dormant while eking out a living in nutrient-poor environments. Much the same pertains to recently discovered seafloor bacteria and single-celled archaea, whose *in situ* (presumably basal) metabolic rates are ~10,000 times slower than lab cultures; at ~$10^2$ erg/s/g, they apparently barely qualify as being alive and may represent an absolute lower limit for life to survive (Roy *et al.* 2012); such extremely low metabolizing microbes have also been found deep below Earth's continents (Lin *et al.* 2006; Danovaro *et al.* 2010). Given the large diversity of cell types, estimates of $\Phi_m$ for realistically metabolizing microbes remain uncertain, although their wide range of values ($10^{2-4}$ erg/s/g) probably approximates those of simple, prokaryotic cells that emerged near the dawn of life.

Microbes and unicells have always been problematic for evolutionary biologists. Darwin largely left them out of his seminal explication of biological evolution, as did the 20$^{th}$-century authors of the Modern Synthesis (Darwin 1859; Mayr 1982). Even so, microbes' genes are still made of DNA, their proteins comprised of the same set of 20 (plus two rare) amino acids, and their energy needs stored in ATP. Microbes are part of the inclusive cosmic-evolutionary scenario, yet here we put them aside in order to focus on normal (healthy) and abnormal (cancerous) cells in human bodies.

**Optimal Range for Human Cells**

Human cells accommodate huge numbers of different molecular components interacting in complex biochemical networks that are not well understood. Here, we consider the cell as a system—a metabolic entity that mainly processes energy, on the whole, like all other complex systems in Nature. Knowledge of metabolic pathways and their multiple interactions is unnecessary when treating cells in bulk; scale-free analysis here is restricted to individual cells interacting with their environment, not concurrently communicating with other cells. (A fuller analysis would incorporate networks, perhaps starting with unicellular yeast, although Nature does display a wealth of diverse phenomena that seem to be scale-free [Mason and Verwoerd 2007; Keller 2005].) All things considered, cells, much like their larger networks, are dynamic steady states that can either survive and flourish (with optimal energy) or degrade badly and terminate (non-optimal energy—*cf.*, Paper I, Chaisson 2014).

Today's eukaryotic, mammalian cells are more massive and richer genomically than prokaryotic, microbial cells that dominated life on early Earth ~3 Gya. (Overlaps and outliers pervade the biological world; some simple bacteria, such as *Epulopiscium*, which thrives in the gut of surgeonfish, are bigger than many complex cells—making $\Phi_m$ somewhat smaller and thus less complex.) Even so, it is unclear how significant $\Phi_m$ values are for individual mammalian cells functioning alone and independently of others within their normal bodily systems. Cells likely operate more efficiently when embedded alongside myriad other cells in a complete living system. Analysis of such individual cells, apart from their parent bodies, might be as futile as that of individual neurons firing separately in a brain or single transistors amplifying in a computer; neither one neural circuit nor one silicon chip comprises a complete, functioning system, and likewise a single mammalian cell hardly constitutes a complex system per se. Remove a neuron from a brain, a chip from a computer, or a cell from a body, and each stops working. Even so, cells and clusters of cells are the focus of cancer research, and so comparisons of $\Phi_m$ values for healthy and cancerous cells might be instructive, especially when measured *in vivo* within their respective bodies; however, $O_2$ consumption rates for separate cell types are very difficult to obtain and are only now becoming feasible with advances in imaging technology.

For resting humans (basal state), thus on-average for their component mammalian cells as noted above, $\Phi_m \approx 11,000$ erg/s/g. This value increases for active states, often by as many as several factors; running, walking, and just sitting upright yield values of 45,000, 28,000, and 15,000 erg/s/g, respectively (Ainsworth 2012). Such active tasks are functionally more complex than lying in bed motionless and they do require more energy flow (per unit mass), hence their enhanced values of $\Phi_m$. If humans acquire, store or express too little energy, or too much, we die; same for a plant if it is insufficiently watered or drowned, and likewise for a star whose energy flow is too little (no longer fuses) or too much (explodes as a supernova). Data for complex systems represented by the curve in Figure 1, from galaxies to society (among myriad systems in between), imply that they all function best within optimal ranges of $\Phi_m$—not too high yet not too low—each type of system having its own range (and not just a single, ideal value) of optimality. Such optimal ranges, wherein complex systems build structure and operate functionally, yet outside of which they terminate, is an important consideration regarding a novel clinical strategy of potential interest to the biomedical community.

**New Attack on Cancer**

Most primary cancer cells are prodigious consumers of glucose, hence seize and maintain higher energy flows than normal cells that rely mainly on $O_2$ uptake (Warburg 1956; Gatenby and Gillies 2004; Hsu and Sabatini 2008; Vander Heiden *et al.* 2009; Koppenol *et al.* 2011); tumors also upregulate glycine during proliferation (Jain *et al.* 2012), implying that cancer cell dynamics and separation are likely interrelated (Enderling *et al.* 2009). Tumors are also widely observed (via direct imagery) to become disproportionately massive while experiencing rapid sigmoidal growth to diameters of ~200 μm (Drasdo and Hohme 2005), and malignant breast tumors known from mammography scans to be ~30% denser than the tissue of origin (Aiello *et al.* 2005). If $\Phi_m$ is a genuine complexity metric, then depressed values of $\Phi_m$ are expected since dysfunctional cancerous cells are less constructively complex, that is more chaotically disordered and entropic, than healthy cells (Seyfried 2012); cancer ravages parts of organisms in which it resides, thus tumors' energy rate densities should decrease with cellular corruption. And that does match what is generally observed in the laboratory, as noted further below.

Figure 4 shows the change in energy rate density as normal and cancerous cells develop and age. Data approximations for single-cell dynamics were taken from *in vivo* (Ramanujan and Herman 2008) and *in vitro* studies (Drasdo and Hohme 2005), as well as from computer modeling (Enderling *et al.* 2009); estimates of $\Phi_m$ were then made accordingly. The figure's graph shows how the value of $\Phi_m$ departs from normality, as expected for a less ordered, more chaotic system; cancer disrupts cellular organization, causing entropy of tumors to rise. Although cancer cells have uniquely high aerobic glycolysis (metabolic rate), a typical tumor's mass grows faster (roughly as the diameter cubed, thus two to as many as four orders of magnitude compared to a single cell, *e.g.*, a typical 20 → 200 μm crossection growth implies ~$10^3$ times mass increase) than its power intake rises (generally one or two orders of magnitude, with variations among many different types of tumors; Warburg 1962; Ramanathan *et al.* 2005; Moreno-Sanchez *et al.* 2007). These are estimates based on a variety of findings reported in the vast medical literature (*e.g.*, inferred from elevated $O_2$ respiration rates and highly active mitochondria in stem cell lines, [Zu and Guppy 2004], magnetic resonance imaging of malignant gliomas [Cao *et al.* 2006], and positron emission tomography applied to oncology [Shields 2006]), and not derived directly from controlled measurements of tumors' power intake and resultant mass as would be preferred. If correct, $\Phi_m$ (*i.e., specific* metabolic rate) decreases as heightened metabolism directs previously normal cells toward (and perhaps outside) lower bounds of optimality, thereby stressing them, sometimes damaging them, and occasionally even destroying them completely. Such lower-than-normal values of $\Phi_m$ are also consistent with the widespread notion that, with fewer and malfunctioning mitochondria present to process glucose, cancer cells resemble primitive organisms, as discussed above for prokaryotic cells and simple microbes (Davies and Lineweaver 2011). Although tumors in mice, rats, and humans, as well as tumor types in various human organisms per se, differ considerably in energy metabolism and growth rate, the *general* trend of decreasing $\Phi_m$ with cancer progression apparently holds for most disadvantaged clinical cases, as inferred from reports throughout the medical literature (Seyfried 2012).

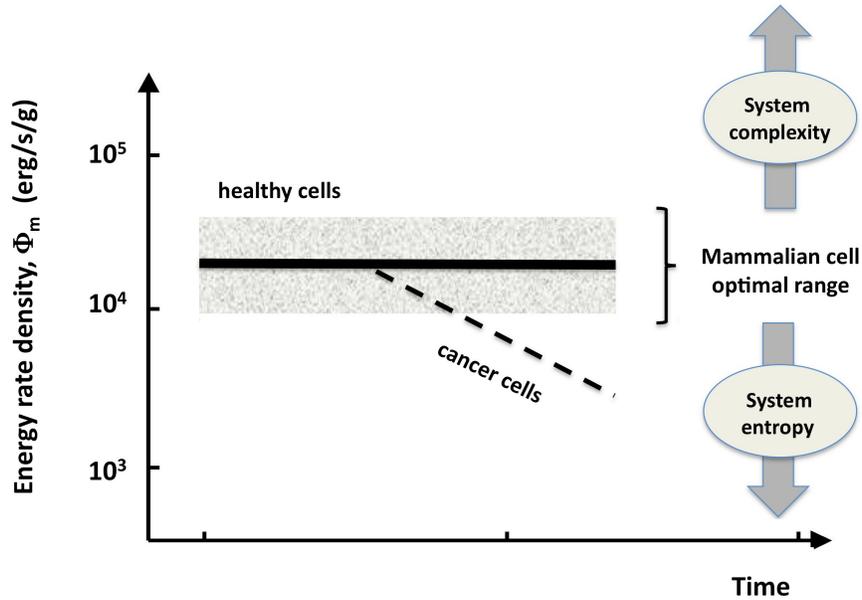

**Figure 4** – *This simple graph is one physicist's way, minus the devilish details, of schematically visualizing cancer development in mammals. Every complex system has a range of optimality for the flow of its energy per unit mass. Outside their nominal range of energy rate density, $\Phi_m$, systems fail to function properly or even to maintain their degree of structural complexity. The graph depicts how the aging of a normal, healthy, mammalian cell (solid horizontal line within the shaded grey, optimal area) can become cancerous when its value of $\Phi_m$ progressively lessens toward death (dashed line projected beyond optimality). Numerical values are unspecified for the temporal axis because cancer cell lines and malignant tumors grow and age so variably* in vivo. *The proposal made in the present study seeks to reverse cancer's decline in $\Phi_m$, namely to raise it back to within the range of optimality and thus to return tumors to better health.*

Normally, most of a cell's energy is used to produce ATP in mitochondria and synthesize macromolecules in tissues. Three aggregates of biomass are often identified by clinicians: Normal, healthy tissue is complex since it is well differentiated—much like a star or planet that complexifies while enhancing thermal and chemical gradients from core to surface. The healthiest cells are often nearly perfectly differentiated and thus not neoplastic (newly abnormal), hence of low entropy. Moderately differentiated neoplasms, which are often benign, uncancerous tumors, have some normality yet some deformity as well, thus are moderately entropic. Poorly differentiated neoplasms, such as malignant tumors, have little of the ordered morphological appearance of normal cells and are highly entropic. That cancer cells seem less complex than normal ones agrees with recent findings that higher molecular network entropy for cancer sites (prostate excepted) correlates with lower probability of 5-year survival (Breitkreutz *et al.*, 2012). Such neoplasms are often regarded as microcosms of clonal evolution within ecological micro-environments, where mutant cells compete for space and resources while evading predation by healthy immune systems (Nowell 1976; Merlo *et al.* 2006). The idea of cancer as an evolutionary problem accords well with the larger scenario of cosmic evolution, which is facilitated by and naturally selecting for optimal energy flows, as discussed throughout these two coupled papers. Cancer therapies might therefore conceivably benefit from evolutionary perspectives.

A hallmark of cancer cells is their enhanced use of energy to feed tumors, whose enzymes and glucose transporters are increased and whose metabolic pathways are over-expressed (Moreno-Sanchez *et al.* 2007); much of cancer is several-fold active, its energy budget considerably ramped up. A prime objective of the medical community seeks to slow or stop tumor growth without affecting normal, nearby tissue, ideally non-invasively. Radiologists disrupt tumors by irradiating them with large doses of energy density, which is why radiotherapy is an effective, yet often collaterally harmful cancer treatment; this method essentially seeks to dramatically increase $\Phi_m$, thereby killing cells by driving $\Phi_m$ above their normal range of optimality. Pharmacologists also destroy cancer by attenuating glycolysis, thus suppressing cancer progression by depriving tumor cells of metabolic energy; such chemotherapy, often accompanied by serious side effects, seeks to substantially decrease $\Phi_m$ by starving cancer cells, thereby killing them by forcing $\Phi_m$ below the cellular range of optimality. Neither method, as well as direct invasive surgery, actually cures cancer as much as tries to destroy it outright (Kroemer and Pouyssegur 2008).

Yet another therapeutic strategy, inspired by cosmic-evolutionary insight, might inhibit growth and proliferation of this dreaded disease by seeking an actual cure. Suppose that, instead of clinically killing them, cancerous cells could revert (or adapt) to normality—that is, increase their $\Phi_m$ values modestly, without themselves dying and without damaging host tissue or organ functionality. In principle, cancer cells' $\Phi_m$ values can be enhanced, thus making them more healthy and less entropic, by either decreasing their tumors' mass or increasing their tumors' energy. The former is the traditional route to eradicate cancer outright. The latter is an alternative method that seeks to heal or at least contain cancer: By feeding cancer cells moderately more energy—neither much more nor any less—their $\Phi_m$ values might rise enough to promote normal cellular health. Such "energy-enriched metabolic intervention" could escalate energy delivery to tumor cells either by increasing their supply of $O_2$ or delivering chemical energy via designed (anti-neoplastic) drugs; modest thermal energy might also help. Hyperbaric chambers that deliver high doses of $O_2$, hence an energy supplement, seem to aid cancer-ridden patients (Tibbles and Edelsberg 1996); even slight heating of tumors, notably by low-energy microwaves, can positively disrupt some cancer cells that exhibit greater thermal sensitivity than normal cells (Cavaliere *et al.* 1967). Moderately elevated (not less) blood flow into tumors might also return some of their carcinogenic cells to pseudo-normality; high-resolution optical imaging in clinical settings shows that such novel anti-cancer therapies often prune and/or remodel abnormal tumor vessels, restoring some of their vascular tissue's structure and function (Jain 2013). Pressure too, which delivers mechanical forces to micro-environments around cancer cells thereby raising temperature and delivering energy where gently applied, might guide malignant cells back toward normal growth patterns; fluorescence imaging shows uncompressed colonies of cancerous cells to be large and disorganized, in contrast to compressed colonies that are smaller and more organized (Venugopalan *et al.* 2012)—much as expected from the above discussion if $\Phi_m$ is a valid complexity metric. What's unknown is whether, by slightly elevating their energy input, tumors will progress faster than their enhanced energy intake, thus continuing to lower their $\Phi_m$ values and damaging their host organisms still more—or whether their rate of energy enhancement could exceed their rate of growth, thus physically returning them to healthier, differentiated neoplasms without toxic chemical side effects and even without biochemically fixing the genetic mutations responsible for malignancy.

**Implications for Cancer Research**

It might seem counterintuitive that a potential cure for energy-hungry cancer cells would entail giving them yet more energy. The alternative hypothesis offered here suggests increasing tumors' $\Phi_m$ by feeding them moderately yet faster than they can grow further, thereby raising the dashed line of Figure 4—or at least causing it to depart from normality less rapidly and less frequently, thus halting tumor progression. Such peculiar reasoning resembles how paradoxically, in economics, the unemployment rate can increase, despite thousands of jobs added each month, when the total workforce also grows at an even faster rate; or in cancer-related human behavior, how smoking can become less popular each year (now ~18% of all living people smoke, down from ~26% a few decades ago), even though the total number of daily smokers still grows globally (with more than a quarter-billion added since 1980; World Health Organization 2013). Relative rate ratios are often revealing.

New and effective therapeutic strategies and treatment regimens aimed to influence metabolic regulation in cancer cells might be realized by targeting those cells with somewhat increased energy, albeit not the excessive energy used in radiotherapy that often adversely affects normal tissue of host cells and also (presumably) not by means of elevated amounts of glucose on which tumors thrive. Targeted delivery by antitumor drugs of moderate doses of additional energy (possibly by upregulating $O_2$) to cancer cells might help normalize them by returning their energy rate densities to the usual range of optimality for healthy cells. Patient survivability could conceivably improve if the promised tools of synthetic biology and bioengineering, aided smartly perhaps by pure and applied physics, not only inhibit cancer-signaling pathways, but especially deliver well adjusted amounts of additional molecular energy throughout cellular networks within and around malignant tumors.

It would be most ironic if the cosmology of cosmic evolution might inform modern medicine regarding its conduct of cancer research today. Surely, the science of biology, upon which medicine is firmly based, would benefit from having a grand quantitative theory, and perhaps cosmic evolution could provide a very broad one, along with a set of underlying principles that guide changes within and among all complex systems, including the birth, life, and death of human systems near and dear to us.

## VI. Summary

I have no illusions regarding the reception of potential applications of cosmic evolution for the health, wealth, and security of humankind, even perhaps for the destiny of modern civilization. Reasons abound why such systems-based practicalities will not likely be embraced by society in general and specialists in particular. Foremost among them, the analytical approach espoused here is well outside mainstream research and development for each of the cultural topics examined in this study. Climate scientists having vested interests in their favorite global-warming models will only reluctantly admit to overlooking a basic thermodynamic ingredient that could well affect long-term meteorological outcomes. Computer engineers who envision today's technological society as a hard, functioning machine obeying information theory will likely reject a tendency for humans and machines to favorably enter into a soft, adaptive symbiosis for the benefit of each. Classical economists will almost surely ignore suggestions that our global economy can be profitably modeled as a non-equilibrium system, with rules, regulations, and inviolable physical laws that inherently guide the growth and organization of cities without excessively degrading environments beyond. And medical oncologists will be slow to welcome clinical strategies focused on metabolomics rather than genomics, thereby attempting to cure cancer, rather than killing it, by actively altering energetic rates of carcinogenesis.

I never imagined that a subject so grand and highbrow as cosmic evolution might have any practical applications. Scientific narratives about origins and evolution are firmly rooted in the past, and, with both chance and necessity engaged, they cannot forecast specific events in the future. Writ large, evolution is indeed unceasing, uncaring, and unpredictable. Even so, the broad concepts and empirical findings of these two coupled papers display some observable trends among many variations; and it is on the basis of those general trends that novel insights emerge regarding the current state and future fate of social systems on Earth. Analyses of the past by means of the interdisciplinary scenario of cosmic evolution, whose roots extend far back into deepest time, can help humanity identify new issues and propose new solutions that might aid our risky trajectory along time's future arrow in ways that *go unnoticed in more specialized, disciplinary science*.

Everlasting evolution and rising complexity may well be hallmarks of Nature, especially given that the Universe expands at an accelerating rate.  All reasonably accords with the known laws of physics, and no new science seems needed to explain, in general terms, the origin and evolution of complex systems as islands of order embedded in wider environments of growing disorder.  The many energy-rate-density curves graphed in these two papers likely continue increasing indefinitely for those systems able to survive by exploiting optimal energy flows, among many other systems that are not so favored and thus succumb to rapid disaster or slow extinction.  Whether civilization endures or not—the choice is probably ours—the stars and galaxies will surely continue shining, twirling, and complexifying, with or without sentient beings on Earth or anywhere else.  Cosmic evolution can help empower human beings in countless ways to understand not only the importance of utilizing the essentially infinite resource of our parent star, but also how well-managed and optimally energized complex systems can practically safeguard the destiny of humankind.

## Acknowledgements

I thank faculty colleagues at the Harvard-Smithsonian Center for Astrophysics and students at Harvard College for many trenchant discussions of potential practical applications of cosmic evolution. Jack Ridge (Tufts Univ.), Dennis Bushnell (NASA Langley), Ed Rietman and Heiko Enderling (St. Elizabeth's Hospital, Boston), and Herman Suit (Harvard Medical School) have been especially helpful.

Blackmore, S., *The Meme Machine*, Oxford Univ. Press, Oxford, 1999.

Bloom, D.E., Canning, D., and Fink, G., Urbanization and the Wealth of Nations, *Science*, 319, 772-775, 2008.

Boulding, K.E., *Ecodynamics: A New Theory of Societal Evolution*, Sage Publications, New York, 1978.

Breitkreutz, D., Hlatky, L. and Rietman, E. and Tuszynski, J.A., Molecular signaling network complexity is correlated with cancer patient survivability, *PNAS*, 109, 9209-9212, 2012.

Brown, J.H., *et al.* [11 authors]], Energetic limits to economic growth, *BioScience*, 61: 19–26, 2011.

Brunner, P.H., Reshaping Urban Metabolism, *J. Industrial Ecology*, 11, 11-13, 2007.

Buchanan, M., *Forecast*, Bloomsbury, New York, 2013.

Budyko, M., The effect of solar radiation variations on the climate of the Earth, *Tellus*, 21, 611-619, 1969.

Bureau of Economic Analysis, *National Economic Accounts*, U.S. Dept. of Commerce, Washington, 2013.

Cao, Y., Sundgren, P.C., Tsien, C.I., Chenevert, T.T. and Junck, L., Physiologic and metabolic magnetic resonance imaging in gliomas, *J. Clinical Oncology*, 24, 1228-1235, 2006.

Cavaliere, R., Ciocatto, E.C., Giovanella, B.C., Heidelberger, C., Johnson, R.O., Margottini, M., Mondovi, B., Moricca, G., Rossi-Fanelli, A., Selective Heat Sensitivity of Cancer Cells, *Cancer*, 20, 1351-1381, 1967.

Chaisson, E.J., Complexity: An Energetics Agenda, *Complexity*, 9, 14-21, 2004.

Chaisson, E.J., Energy, Ethics, and the Far Future, *Energy Challenges: The Next 1000 Years*, Foundation for the Future Proceedings, pp 131-138, Seattle, 2007.

Chaisson, E.J., Long-term Global Heating from Energy Usage, *Eos Transactions of the American Geophysical Union*, 89, 253-254, 2008.

Chaisson, E.J., Energy Rate Density as a Complexity Metric and Evolutionary Driver, *Complexity*, 16, 27-40, 2011.

Chaisson, E.J., A Singular Universe of Many Singularities: Cultural Evolution in a Cosmic Context, in *Singularity Hypotheses*, Eden, A.H., *et al.* (eds.), pp 413-438, Springer, Berlin, 2012.

Chaisson, E.J., Using complexity science to search for unity in the natural sciences, in *Complexity and the Arrow of Time*, Lineweaver, C.H., Davies, P.C.W. and Ruse, M. (eds.), pp 68-79, Cambridge Univ. Press, Cambridge, 2013.

Chaisson, E.J., *Cosmic Evolution I: The Natural Science Underlying Big History*, in press, 2014.

Chaisson, E. and McMillan, S., *Astronomy Today*, 8th ed., Pearson, San Francisco, London, 2014.

Christian, D., The role of energy in civilization. *J. World History*, 14, 4-11, 2003.

Cleveland, C.J., Costanza, R., Hall, C.A.S. and Kaufmann, R., Energy and the U.S. Economy: A Biophysical Perspective, *Science*, 225, 890-897, 1984.

Cook, E., *Man, Energy, and Society*, W.H. Freeman, San Francisco, 1976.

Costanza, R., Cumberland, J.C., Daly, H.E., Goodland, R. and Norgaard, R., *An Introduction to Ecological Economics*, St. Lucie Press, Baca Raton, 2009.

Dang, C.V., Links between metabolism and cancer, *Genes & Dev.* 26, 877-890, 2012.
30

Lazar, M.A. and Birnbaum, M.J., De-Meaning of Metabolism, *Science*, 336, 1651-52, 2012.

Lancaster, K., *Mathematical Economics*, Macmillan, New York, 1968.

Lehninger, A.L., Nelson, D.L. and Cox, M.M., *Principles of Biochemistry*, Worth Publishers, New York, 2nd ed, 1993.

Leontief, W.W., *Input-Output Economics*, Oxford Univ. Press, Oxford, 1966.

Levine, A.J. and Puzio-Kuter, A.M., The Control of the Metabolic Switch in Cancers by Oncogenes and Tumor Suppressor Genes, *Science, 330,* 1340-1344, 2010.

Lin, L-H, *et al.*, Long-term sustainability of a high-energy, low-diversity crustal biome, *Science*, 314, 479-482, 2006.

Madsen, J.G., Wang, T., Beedholm, K., and Madsen, P.T., Detecting spring after a long winter. *Biology Letters*, 9, 20130602, 1-5, 2013.

Malthus, T.R., *An Essay on the Principle of Population*, J. Johnson, London, 1798.

Mankiw, N.G., *Macroeconomics*, 7th ed., Worth, New York, 2010.

Moavenzadeh F., Hanaki K. and Baccini P., *Future Cities: Dynamics and Sustainability,* Kluwer Academic, Amsterdam, 2002.

Marglin, S.A., *The Dismal Science*: *How Thinking Like an Economist Undermines Community*, Harvard Univ. Press, Cambridge, 2008.

Marshall, E., Cancer research and the $90 billion metaphor. *Science,* 331, 1540–1541, 2011.

Mason, O. and Verwoerd M., Graph theory and networks in biology, *IET Syst. Biol.*, 1, 89–119, 2007.

Mayr, E., *The Growth of Biological Thought*, Harvard Univ. Press, Cambridge, 1982.

Merlo, L.M.F., Pepper, J.W., Reid, B.J. and Maley, C.C., Cancer as an evolutionary and ecological process, *Nature Cancer Reviews*, 6, 924-935, 2006.

Modelski, G., *World Cities*, Faros, Washington, 2003.

Modis, T., Why the Singularity Cannot Happen, in *Singularity Hypotheses*, Eden, A.H., *et al.* (eds.), pp 311-339, Springer, Berlin, 2012.

Monod, J., *Chance and Necessity*, Knopf, New York, 1971.

Moore, G.E., Cramming more components onto integrated circuits, *Electronics*, v38, no.8, 1965.

Moreno-Sanchez, R., Rodriguez-Enriquez, S., Marin-Hernandez, A. and Saavedra, E., Energy metabolism in tumor cells, *FEBS Journal*, 274, 1393-1418, 2007.

Motter, A. and Campbell, D., Chaos at fifty, *Physics Today*, 66, 27-35, 2013.

Mumford, L., *The Culture of Cities*, Harcourt Brace, New York, 1970.

Nowell, P.C., The clonal evolution of tumor cell populations, *Science*, 194, 23-28, 1976.
33